# Prediction of Radiation Fog by DNA Computing

KUMAR S. RAY
Electronics and Communication Science Unit
Indian Statistical Institute
203, B.T Road, Kolkata-700108, India
E-mail: ksray@isical.ac.in
Tel: +91 8981074174

MANDRITA MONDAL
Electronics and Communication Science Unit
Indian Statistical Institute
203, B.T Road, Kolkata-700108, India
E-mail: mandrita_1984@yahoo.co.in
Tel: +91 9830354798

**Abstract**
In this paper we propose a wet lab algorithm for prediction of radiation fog by DNA computing. The concept of DNA computing is essentially exploited for generating the classifier algorithm in the wet lab. The classifier is based on a new concept of similarity based fuzzy reasoning suitable for wet lab implementation. This new concept of similarity based fuzzy reasoning is different from conventional approach to fuzzy reasoning based on similarity measure and also replaces the logical aspect of classical fuzzy reasoning by DNA chemistry. Thus, we add a new dimension to existing forms of fuzzy reasoning by bringing it down to nanoscale. We exploit the concept of massive parallelism of DNA computing by designing this new classifier in the wet lab. This newly designed classifier is very much generalized in nature and apart from prediction of radiation fog this methodology can be applied to other types of data also. To achieve our goal we first fuzzify the given observed parameters in a form of synthetic DNA sequence which is called fuzzy DNA and which handles the vague concept of human reasoning.

## 1. Introduction

In this paper, we attempt to predict radiation fog by similarity based fuzzy reasoning using synthetic fuzzy DNA. We replace the logical aspect of fuzzy reasoning by DNA chemistry. Thus the tedious choice of a suitable implication operator to translate a rule to a fuzzy relation matrix necessary for conventional fuzzy reasoning can easily be avoided. Our DNA computing model is based on double stranded DNA sequences, whereas, most of the existing DNA computing models as stated above are based on the single stranded DNA sequences. In our DNA computing model we choose double stranded DNA molecules for a specific aim of measuring similarity between two DNA sequences. The similarity measure between two DNA sequences is essential for realizing the similarity based fuzzy reasoning in the wet lab algorithm. In the present approach, whenever we talk about DNA sequences, we essentially mean double stranded fuzzy DNA sequences. For fuzzification of synthetic double stranded DNA sequence, we have used another sequence representing the membership value of the double stranded DNA sequence. Note that, we have developed a completely new measure of similarity based on base pair difference which is absolutely different from the existing measure of similarity (Yeung and Tsang, 1997; 1998) and which is very much suitable for similarity based fuzzy reasoning using DNA computing. Thus, in the present approach in one hand we overcome the existing drawbacks of fuzzy reasoning based on fuzzy logic (Zadeh, 1970) and in other hand we adopt a new measure of similarity which is different from existing measure of similarity and which is very much suitable wet lab implementation of the similarity based fuzzy reasoning. Further in present wet lab procedure for prediction of radiation fog we can exploit massive parallelism of DNA computing. One essential difference between the present model of DNA computing and the existing models as stated above is that, the ultimate result of the wet lab of the present method provides the multi valued status which is not strictly yes or no; but, always something in between (in terms of grade of membership value represented by DNA sequences which are attached with another DNA sequence of the inferred consequence). Such flexibility of the present model of DNA computing is achieved using fuzzy DNA. In this paper we have to predict the possibility of visibility based on the given parameters (i.e. dew point, dew point spread, the rate of change of dew point spread per day, wind speed and sky condition) of radiation fog. The present method of classification in the wet lab is so generalized that it can be applied to other types of data and basically it is an expert system approach to classifier design where intuition and experience, which is essentially derived from the perception of an expert, can be easily incorporated to handle real life problems.

## 2. Fuzzification of DNA Sequence

Usually a set becomes fuzzy when each element of the set is attached with a membership value which indicates the possibility of belongingness of the element to the set. In this paper, for wet lab DNA computing, a ten bases long DNA sequence is treated as an element of a discrete universe (see tables 2 to 7) over which several primary fuzzy sets are defined. Hence for fuzzification of a ten bases long DNA sequence we attach some membership value to it. But we are essentially concerned with DNA computing in the wet lab. Therefore, the numerical membership value is also codified into DNA sequence of variable length (see table 1 and 2) depending upon different numerical values of membership. Thus, for fuzzification of ten bases long DNA sequence, a particular variable length of DNA sequence is attached with a particular ten bases long DNA sequence to indicate its membership value. As we always deal with double stranded DNA sequences, each ten base DNA sequence is coupled with its complementary pair. Similarly each variable length DNA sequence representing the membership value of ten base DNA sequence is also coupled with its complementary pair. Here we consider two types of membership values (see table 1). Each membership value of each table is represented by variable length DNA sequences as shown in the right hand side column of each table. These choices of membership value are not fixed. Depending on the need of design we may have other types of representation of membership value. The membership value of the type shown in table 1 is considered to represent the primary fuzzy sets of the antecedent parts and consequent parts of fuzzy IF-THEN rules (see section 5.1).

Table 1. Representation of the membership value by variable length DNA sequence

| Membership value | DNA sequence (5' to 3') |
| --- | --- |
| 0.1 | AT |
| 0.2 | ATCC |
| 0.3 | ATCCAT |
| 0.4 | ATCCATGC |
| 0.5 | ATCCATGCCG |
| 0.6 | ATCCATGCCGAT |
| 0.7 | ATCCATGCCGATCC |
| 0.8 | ATCCATGCCGATCCAT |
| 0.9 | ATCCATGCCGATCCATGC |
| 1.0 | ATCCATGCCGATCCATGCCG |

**2.1 Fuzzy DNA**

We introduce the concept of fuzzy DNA as follows;

Let, our universe of discourse be discrete. Let an arbitrary ten base DNA sequence be a basic element of discrete universe (refer table 2 to 7). Thus, several elements of discrete universe are represented by several ten base DNA sequences which are arbitrarily chosen. For instance, the universe dew point (see table 2) is discretized (quantized) as shown by the left most column of table 2 and each discrete element of the quantized universe of dew point is represented by ten base DNA sequence, i.e. the discrete element $-30 \leq Td < -25 \equiv$ CTGGACTGGA etc. The range of a discrete domain is represented by different ten bases long DNA oligonucleotide (table 2 to 7). Over each range of discrete domain we have well defined primary sets like very low, low, medium etc. (refer table 7). A ten base DNA sequence (coupled with its complementary pair), taken from any table2 to 6, attached with an appropriate DNA sequence of variable length (coupled with its complementary pair) taken from table 1, together form a fuzzy DNA. For example,

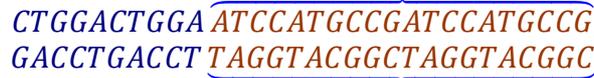

*CTGGACTGGA* *ATCCATGCCGATCCATGCCG*
*GACCTGACCT* *TAGGTACGGCTAGGTACGGC*

is a fuzzy DNA. For wet lab computation a string of fuzzy DNA sequence which consists of several fuzzy DNA strands, represents a primary fuzzy set over a given discrete universe. Note that, whenever we talk about DNA sequences we always consider double stranded DNA sequences. Such double stranded DNA sequence is necessary to perform a specific task of similarity measurement between two primary fuzzy sets in our wet lab computation which will be discussed later (in section 5.3). Therefore, a primary fuzzy set, say, 'dry' defined over the discrete universe of dew point (refer table 2) is represented as,

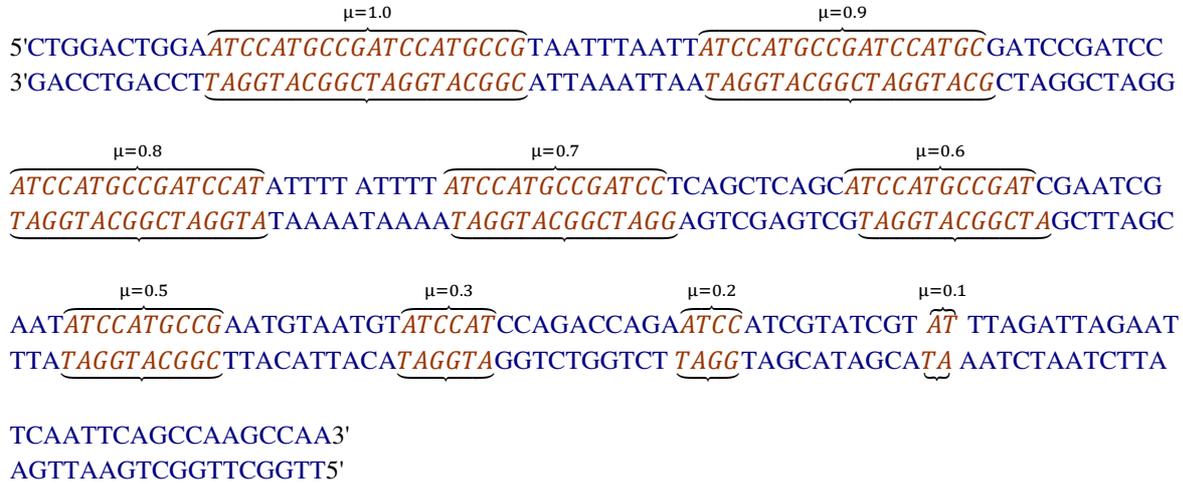

TCAATTCAGCCAAGCCAA3'
AGTTAAGTCGGTTCGGTT5'

Figure 1: DNA sequence representing 'dry' of discrete universe dew point

Similarly, we can represent other primary fuzzy sets as shown in table 2 to 7 in terms of double stranded fuzzy DNA sequences. For the present classification problem we have one class i.e. possibility of visibility which has five primary fuzzy sets as shown in table 7.

The existing models of DNA computing proposed by L. Adleman and R. J. Lipton are based on single stranded DNA molecules. But, in our paper the model deals with double stranded fuzzy DNA molecules. Further our model of DNA-computing can handle vague and uncertain environment which is not possible by other approaches to DNA-computing as stated above.

**3. Semantics of DNA computing**

The linguistic interpretation of the end result of the wet lab provides the semantic of our present model of computation. If the linguistic interpretation of the end result of the wet lab is acceptable to the commonsense of a human being we would consider the interpretation to be true/ partially true/ more or less true etc. depending upon the degree of acceptability to the commonsense of a human being; otherwise it becomes false/nearly false etc. in the same manner. For instance, in section 5.1 the fuzzy DNA sequence (see the first row of table 2 to 7) of the consequence of rule 1, gives the possibility of visibility is linguistically interpreted as very high i.e. true. Like the same way the linguistic interpretation of the end result is 'low' i.e. more or less true. Such linguistic interpretation of the end result of the wet lab experiment is perceptually acceptable to our commonsense and treated as valid. Hence this particular linguistic interpretation is treated as the semantic of our DNA computation.

As our ultimate goal of this model of computation is to develop a fundamental notion of computation with perception we choose this particular approach to semantic which essentially tries to map human perception in a multi valued sense.

## 4. Fuzzy Reasoning System for Expert System Design

The basic concept of fuzzy reasoning was first introduced by Zadeh (Zadeh, 1970). Subsequently extended fuzzy reasoning was developed by Mizumoto (Mizumoto, 1985a, 1985b). We consider the following extended form of fuzzy reasoning system and replace its logical aspect (see appendix A) by using DNA chemistry. A logical aspect of fuzzy reasoning means the particular interpretation which is considered to translate a fuzzy conditional statement to a fuzzy relational matrix (see appendix A) which is further used for compositional rule for inference. Such a choice of particular interpretation to translate a conditional statement to fuzzy relational matrix is completely arbitrary for a given problem. Given a problem for translating a fuzzy conditional statement to a fuzzy relational matrix, we may either go by Mamdani's min operator or by Zadeh's arithmetic rule etc. (see appendix A). To avoid such tedious choice of a suitable implication operator we go by DNA chemistry which means the standard DNA operations and the new similarity measure for wet lab implementation of fuzzy reasoning system. A generalized form of fuzzy reasoning system, as shown below, is several fuzzy conditional propositions combined with else.

That is,
Premise 1 : If $X$ is $A_1$ and $Y$ is $B_1$ then $Z$ is $C_1$ else
Premise 2 : If $X$ is $A_2$ and $Y$ is $B_2$ then $Z$ is $C_2$ else
.
.
Premise $n$ : If $X$ is $A_n$ and $Y$ is $B_n$ then $Z$ is $C_n$ else
Premise $n+1$ : If $X$ is $A'$ and $Y$ is $B'$
___________________________________________
Consequence: $Z$ is $C'$.

We interpret else as union (∪) which is valid for the fuzzy implication $R_c$, $R_p$, $R_{bp}$ and $R_{dp}$ (see appendix A of [Ray and Mondal, 2009]) in above discussed fuzzy implications. Thus we can deduce the consequences $C'$ as,

$$C' = (A' \text{ and } B') \circ [((A_1 \text{ and } B_1) \rightarrow C_1) \cup \ldots \cup ((A_n \text{ and } B_n) \rightarrow C_n]$$
$$= [(A' \circ A_1 \rightarrow C_1) \cap (B' \circ B_1 \rightarrow C_1)] \cup \ldots \cup [(A' \circ A_n \rightarrow C_n) \cap (B' \circ B_n \rightarrow C_n)]$$
$$= C'_1 \cup C'_2 \cup \ldots \cup C'_n \qquad (1)$$

whereas, for the fuzzy implications $R_a$, $R_m$, $R^*$, $R_\#$ and $R_\Delta$ (see appendix A of [Ray and Mondal, 2009]) in above discussed fuzzy implications, else is interpreted as intersection (∩). Thus, the consequences $C'$ for these fuzzy implications are defined as,

$$C' = (A' \text{ and } B') \circ [((A_1 \text{ and } B_1) \rightarrow C_1) \cap \ldots \cap ((A_n \text{ and } B_n) \rightarrow C_n]$$
$$= [(A' \circ A_1 \rightarrow C_1) \cup (B' \circ B_1 \rightarrow C_1)] \cap \ldots \cap [(A' \circ A_n \rightarrow C_n) \cup (B' \circ B_n \rightarrow C_n)]$$

It is noted that the consequence $C'$ is not equal to but contained in the intersection of fuzzy inference results $[(A' \circ A_i \rightarrow C_i) \cup (B' \circ B_i \rightarrow C_i)] \; \forall i$. However, for simplicity of calculation $C'$ will be represented as,

$$= C'_1 \cap C'_2 \cap \ldots \cap C'_n \qquad (2)$$

## 5. Prediction of Radiation Fog by Fuzzy Reasoning System Using DNA Computing

We consider the prediction of radiation fog as a particular application of fuzzy reasoning system. The present result of fuzzy reasoning system based on DNA computing is very much generalized in nature and can be applied to any other pattern classification problem.

For prediction of radiation fog we consider five parameters of radiation fog. These parameters are considered as discrete domains. The domains are as follows:

**i. Dew Point:** The temperature to which humid air can be cooled at constant pressure without causing condensation is called the dew point temperature or dew point. It is represented by Td and measured in (ºC). Here domain of dew point is Td = {-300ºC to 300ºC}. Primary fuzzy sets defined over the said domain are dry, moderate, moist and very moist. The fuzzy sets and their membership functions are given in table 2.

Table 2. Representation of primary fuzzy sets over the discrete domain of Dew Point with membership function

| Td | DNA sequence (5' to 3') | dry | moderate | moist | very moist |
|---|---|---|---|---|---|
| $-30 \leq Td < -25$ | CTGGACTGGA | 1.0 | 0.5 | 0.2 | 0 |
| $-25 \leq Td < -20$ | TAATTTAATT | 0.9 | 0.6 | 0.3 | 0 |
| $-20 \leq Td < -15$ | GATCCGATCC | 0.8 | 0.7 | 0.5 | 0 |
| $-15 \leq Td < -10$ | ATTTTATTTT | 0.7 | 0.8 | 0.6 | 0.1 |
| $-10 \leq Td < -5$ | TCAGCTCAGC | 0.6 | 0.9 | 0.7 | 0.2 |
| $-5 \leq Td < 0$ | CGAATCGAAT | 0.5 | 1.0 | 0.8 | 0.3 |
| $0 \leq Td < 5$ | AATGTAATGT | 0.3 | 0.9 | 0.9 | 0.5 |
| $5 \leq Td < 10$ | CCAGACCAGA | 0.2 | 0.8 | 1.0 | 0.6 |
| $10 \leq Td < 15$ | ATCGTATCGT | 0.1 | 0.7 | 0.9 | 0.7 |
| $15 \leq Td < 20$ | TTAGATTAGA | 0 | 0.6 | 0.8 | 0.8 |
| $20 \leq Td < 25$ | ATTCAATTCA | 0 | 0.5 | 0.7 | 0.9 |
| $25 \leq Td \leq 30$ | GCCAAGCCAA | 0 | 0.3 | 0.6 | 1.0 |

**ii. Dew Point Spread:** The difference between the air temperature (T) and dew point (Td) is termed as dew point spread. It is represented by ΔT and measured in (ºC). Here, the domain of dew point spread is ΔT = {-120ºC to 120ºC}. Primary fuzzy sets defined over the said domain are very saturated, saturated and unsaturated. The fuzzy sets and their membership functions are given in table 3.

Table 3. Representation of primary fuzzy sets over the discrete domain of Dew Point Spread with membership function

| ΔT=T-Td | DNA sequence (5' to 3') | very saturated | saturated | unsaturated |
|---|---|---|---|---|
| $-12 \leq \Delta T < -10$ | TTCGTTTCGT | 1.0 | 0.3 | 0 |
| $-10 \leq \Delta T < -8$ | CAAACCAAAC | 0.9 | 0.5 | 0 |
| $-8 \leq \Delta T < -6$ | CGGAACGGAA | 0.8 | 0.6 | 0 |
| $-6 \leq \Delta T < -4$ | ATCCGATCCG | 0.7 | 0.7 | 0.1 |
| $-4 \leq \Delta T < -2$ | GAATGGAAT | 0.6 | 0.8 | 0.2 |
| $-2 \leq \Delta T < 0$ | GTAGCGTAGC | 0.5 | 0.9 | 0.3 |
| $0 \leq \Delta T < 2$ | ATCCCATCCC | 0.3 | 1.0 | 0.5 |
| $2 \leq \Delta T < 4$ | TAGGATAGGA | 0.2 | 0.9 | 0.6 |
| 4 ≤ ΔT < 6 | CTAAGCTAAG | 0.1 | 0.8 | 0.7 |
| 6 ≤ ΔT < 8 | AGGAAAGGAA | 0 | 0.7 | 0.8 |
| 8 ≤ ΔT < 10 | TAGCTTAGCT | 0 | 0.6 | 0.9 |
| 10 ≤ ΔT ≤ 12 | GCGCGGCGCG | 0 | 0.5 | 1.0 |

**iii. The Rate of Change of Dew Point Spread per Day:** The difference between the dew point spreads of two consecutive days is defined as the rate of the change of spread per day. It is represented by ΔT' and measured in (ºC). Here domain of ΔT' = {-50ºC to 60ºC}. Primary fuzzy sets defined over the said domain are dying and saturating. The fuzzy sets defined over their membership functions are given in table 4.

Table 4. Representation of primary fuzzy sets over the discrete domain of Rate of Change of Dew Point Spread per Day with membership function

| ΔT' | DNA sequence (5' to 3') | saturating | drying |
|---|---|---|---|
| -5 ≤ ΔT' < -4 | GTAACGTAAC | 1.0 | 0 |
| -4 ≤ ΔT' < -3 | AAATAAAATA | 0.9 | 0 |
| -3 ≤ ΔT' < -2 | TGCGATGATT | 0.8 | 0.1 |
| -2 ≤ ΔT' < -1 | CGGAATTAAC | 0.7 | 0.2 |
| -1 ≤ ΔT' < 0 | GTAATCTATG | 0.6 | 0.3 |
| 0 ≤ ΔT' < 1 | GTTAACGGTA | 0.5 | 0.5 |
| 1 ≤ ΔT' < 2 | AGTAGCGTAG | 0.3 | 0.65 |
| 2 ≤ ΔT' < 3 | TACCGTAACT | 0.2 | 0.7 |
| 3 ≤ ΔT' < 4 | AGCGTAGCTA | 0.1 | 0.8 |
| 4 ≤ ΔT' < 5 | CAGTACCGAT | 0 | 0.9 |
| 5 ≤ ΔT' ≤ 6 | TACGCAGGAA | 0 | 1.0 |

**iv. Wind Speed:** Wind speed is the speed of wind in Knts/hr. W represents it. Here domain of wind speed is W = {-5Knts/hr to 25Knts/hr}. Primary fuzzy sets defined over the said domain are too light, excellent and too strong. The fuzzy sets and their membership functions are given in Table 5.

Table 5. Representation of primary fuzzy sets over the discrete domain of Wind Speed with membership function

| W | DNA sequence (5' to 3') | too light | excellent | too strong |
|---|---|---|---|---|
| $-5 \leq W < -2.5$ | GCTATGGCTG | 1.0 | 0.5 | 0 |
| $-2.5 \leq W < 0$ | ATACCGATAG | 0.9 | 0.6 | 0 |
| $0 \leq W < 2.5$ | TAACCCGTGG | 0.8 | 0.7 | 0 |
| $2.5 \leq W < 5$ | CAGTAAGTCT | 0.7 | 0.8 | 0.1 |
| $5 \leq W < 7.5$ | ATTCCGATGC | 0.6 | 0.9 | 0.2 |
| $7.5 \leq W < 10$ | TACAGTACGG | 0.5 | 1.0 | 0.3 |
| $10 \leq W < 12.5$ | GTGCCAGTCT | 0.3 | 0.9 | 0.5 |
| $12.5 \leq W < 15$ | ATGTAACGTG | 0.2 | 0.8 | 0.6 |
| $15 \leq W < 17.5$ | GCAGCTATAC | 0.1 | 0.7 | 0.7 |
| $17.5 \leq W < 20$ | ATGATCGGTA | 0 | 0.6 | 0.8 |
| $20 \leq W < 22.5$ | TAGCCGATGG | 0 | 0.5 | 0.9 |
| $22.5 \leq W \leq 25$ | GTAGATGGTC | 0 | 0.3 | 1.0 |

**v. Sky Condition:** Sky coverage is in terms of percentage of cloud coverage perceptually judged by inspection. It is represented by S and measured in (%). Here domain of S={0% to 100%}. Primary fuzzy sets defined over the said domain is clear sky, partially cloudy and cloudy. The fuzzy sets and their membership functions are given in table 6.

Table 6. Representation of primary fuzzy sets over the discrete domain of Sky Condition with membership function

| S | DNA sequence (5' to 3') | clear | partially cloudy | cloudy |
|---|---|---|---|---|
| $0 \leq S < 10$ | TCTCTATGGA | 1.0 | 0.6 | 0 |
| $10 \leq S < 20$ | CGATAGCAAT | 0.9 | 0.7 | 0.1 |
| $20 \leq S < 30$ | ATGCAATCTT | 0.8 | 0.8 | 0.2 |
| $30 \leq S < 40$ | CGATGACTAC | 0.7 | 0.9 | 0.3 |
| $40 \leq S < 50$ | TCAGACTATG | 0.6 | 1.0 | 0.5 |
| $50 \leq S < 60$ | CGTAACGATT | 0.5 | 0.9 | 0.6 |
| $60 \leq S < 70$ | CGATAGCCAG | 0.3 | 0.8 | 0.7 |
| $70 \leq S < 80$ | ATGCGTAGCG | 0.2 | 0.7 | 0.8 |
| $80 \leq S < 90$ | TACCGGCCGG | 0.1 | 0.6 | 0.9 |
| $90 \leq S \leq 100$ | TACCATAAGA | 0 | 0.5 | 1.0 |

For prediction of radiation fog we have to predict the possibility of visibility. This is also treated as a discrete domain.

**Possibilty of Visibility:** As no standard as well established visibility versus fog (haze) classification exists we consider the following ranges of visibility from international definition of fog (V<1 km.). Depending upon the different ranges of visibility, we consider five primary fuzzy sets as shown below.

Table 7. Representation of primary fuzzy sets over the discrete domain of Visibility with membership function

| visibility V(km.) | DNA sequence (5' to 3') | very low | low | medium | high | very high |
|---|---|---|---|---|---|---|
| V < 1 | CTACCATCAG | 1.0 | 0.7 | 0.5 | 0.3 | 0.1 |
| 1 ≤ V < 5 | TCGGTGATGG | 0.7 | 1.0 | 0.7 | 0.5 | 0.3 |
| 5 ≤ V < 10 | GAGCATAGGG | 0.5 | 0.7 | 1.0 | 0.7 | 0.5 |
| 10 ≤ V < 16 | CTAACTTGTA | 0.3 | 0.5 | 0.7 | 1.0 | 0.7 |
| V > 16 | GTCTGATACG | 0.1 | 0.3 | 0.5 | 0.7 | 1.0 |

Thus, the generalized form of fuzzy reasoning system as stated in section 4 is now reduced to a specific form of fuzzy reasoning system having a finite number of rules and each rule is having five antecedent clauses, consists of dew point, dew point spread, rate of change of dew point spread per day, wind speed sky condition and one consequent clause, consists of possibility of visibility. Without any lack of generality, above mentioned form of fuzzy reasoning system can be extended to any suitable form of fuzzy reasoning system as stated in section 4. The quantized ranges of the antecedent clause (i.e. Td, ΔT, ΔT', W and S) are represented by DNA oligonucleotide sequences in table 2 to 6. The membership function (μ) of different primary fuzzy sets defined over the said quantized domains are also represented in table 2 to 6. The membership values are coded in the form of DNA sequence (table 1). The consequent clause of the radiation fog are represented in table 7. For further details on radiation fog interested readers are referred to appendix C.

### 5.1. Generation of the if-then rules
Here the number of possible rules = $^4C_1 * {}^2C_1 * {}^2C_1 * {}^3C_1 * {}^3C_1$ = 216. The general formula of determining the number of possible rules is the product of the number of primary sets taking one at a time from each of them. The antecedents of 10 random applicable rules from all of the rules along with their corresponding consequents are given in the table 8. These are according to the perception of an expert.

Table 8. Expert rules for prediction of radiation fog

| Td | ΔT | ΔT' | W | S | V(Consequence) |
|---|---|---|---|---|---|
| dry | very saturated | drying | too light | cloudy | very high |
| dry | saturated | saturating | excellent | clear | medium |
| moderate | unsaturated | drying | too strong | partially cloudy | very high |
| moist | saturated | drying | excellent | clear | low |
| moist | unsaturated | saturating | too strong | cloudy | medium |
| very moist | very saturated | saturating | too light | clear | low |
| very moist | saturated | drying | too strong | partially cloudy | very high |
| moderate | saturated | saturating | excellent | cloudy | high |
| moderate | very saturated | drying | too light | clear | medium |
| moderate | very saturated | saturating | excellent | clear | very low |

The first rule is coded in the form of double stranded DNA sequences.

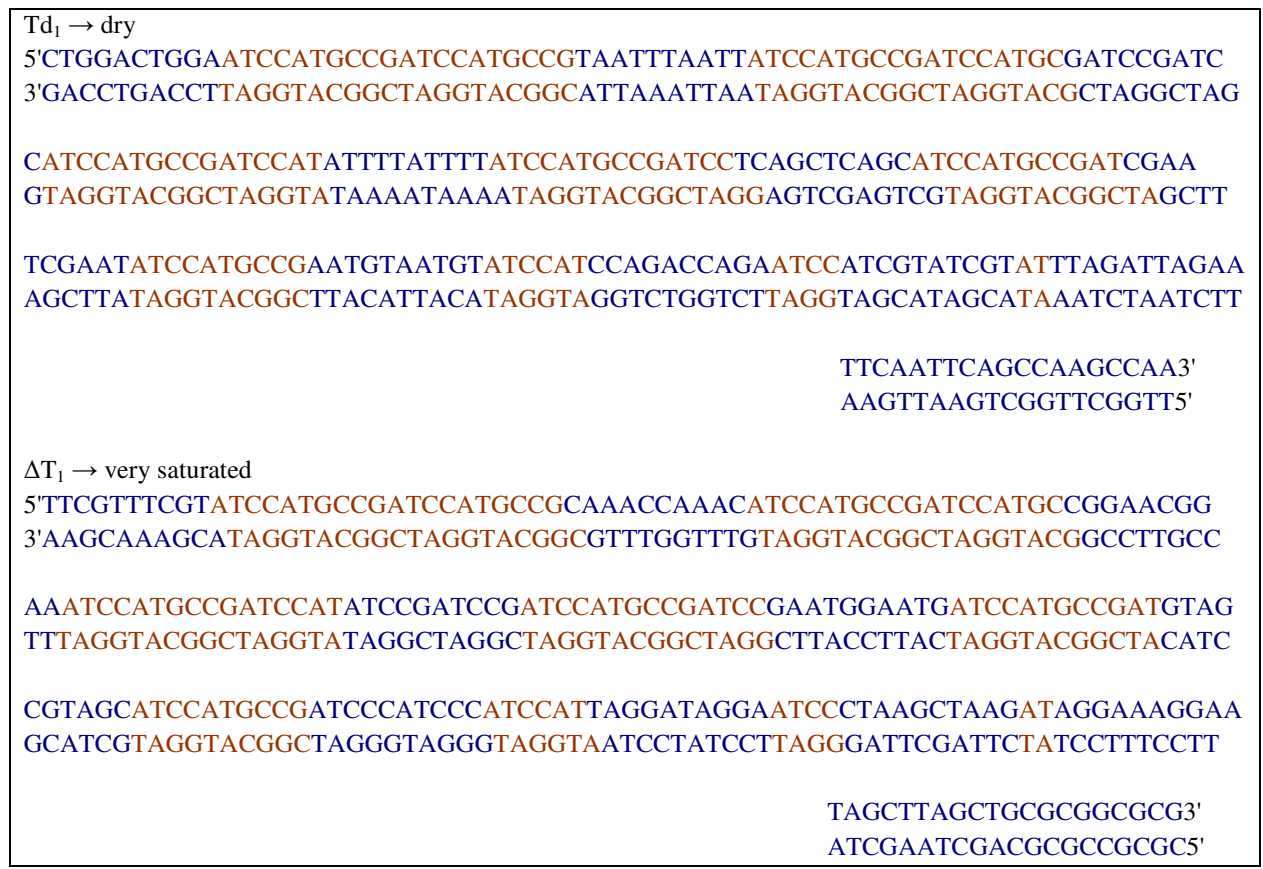

Td$_1$ → dry
5'CTGGACTGGAATCCATGCCGATCCATGCCGTAATTTAATTATCCATGCCGATCCATGCGATCCGATC
3'GACCTGACCTTAGGTACGGCTAGGTACGGCATTAAATTAATAGGTACGGCTAGGTACGCTAGGCTAG

CATCCATGCCGATCCATATTTTATTTTATCCATGCCGATCCTCAGCTCAGCATCCATGCCGATCGAA
GTAGGTACGGCTAGGTATAAAATAAAATAGGTACGGCTAGGAGTCGAGTCGTAGGTACGGCTAGCTT

TCGAATATCCATGCCGAATGTAATGTATCCATCCAGACCAGAATCCATCGTATCGTATTTAGATTAGAA
AGCTTATAGGTACGGCTTACATTACATAGGTAGGTCTGGTCTTAGGTAGCATAGCATAAATCTAATCTT

                       TTCAATTCAGCCAAGCCAA3'
                       AAGTTAAGTCGGTTCGGTT5'

ΔT$_1$ → very saturated
5'TTCGTTTCGTATCCATGCCGATCCATGCCGCAAACCAAACATCCATGCCGATCCATGCCGGAACGG
3'AAGCAAAGCATAGGTACGGCTAGGTACGGCGTTTGGTTTGTAGGTACGGCTAGGTACGGCCTTGCC

AAATCCATGCCGATCCATATCCGATCCGATCCATGCCGATCCGAATGGAATGATCCATGCCGATGTAG
TTTAGGTACGGCTAGGTATAGGCTAGGCTAGGTACGGCTAGGCTTACCTTACTAGGTACGGCTACATC

CGTAGCATCCATGCCGATCCCATCCCATCCATTAGGATAGGAATCCCTAAGCTAAGATAGGAAAGGAA
GCATCGTAGGTACGGCTAGGGTAGGGTAGGTAATCCTATCCTTAGGGATTCGATTCTATCCTTTCCTT

                       TAGCTTAGCTGCGCGGCGCG3'
                       ATCGAATCGACGCGCCGCGC5'

ΔT₁' → drying
5'GTAACGTAACAAATAAAATATGCGATGATTATCGGAATTAACATCCGTAATCTATGATCCATGTTAA
3'CATTGCATTGTTTATTTTATACGCTACTAATAGCCTTAATTGTAGGCATTAGATACTAGGTACAATT

CGGTAATCCATGCCGAGTAGCGTAGATCCATGCCGATTACCGTAACTATCCATGCCGATCCAGCGTAG
GCCATTAGGTACGGCTCATCGCATCTAGGTACGGCTAATGGCATTGATAGGTACGGCTAGGTCGCATC

CTAATCCATGCCGATCCATCAGTACCGATATCCATGCCGATCCATGCTACGCAGGAAATCCATGCCGA
GATTAGGTACGGCTAGGTAGTCATGGCTATAGGTACGGCTAGGTACGATGCGTCCTTTAGGTACGGCT

                                                               TCCATGCCG3'
                                                               AGGTACGGC5'

W₁ → too light
5'GCTATGGCTGATCCATGCCGATCCATGCCGATACCGATAGATCCATGCCGATCCATGCTAACCCGTG
3'CGATACCGACTAGGTACGGCTAGGTACGGCTATGGCTATCTAGGTACGGCTAGGTACGATTGGGCAC

GATCCATGCCGATCCATCAGTAAGTCTATCCATGCCGATCCATTCCGATGCATCCATGCCGATTACAG
CTAGGTACGGCTAGGTAGTCATTCAGATAGGTACGGCTAGGTAAGGCTACGTAGGTACGGCTAATGTC

TACGGATCCATGCCGGTGCCAGTCTATCCATATGTAACGTGATCCGCAGCTATACATATGATCGGTAT
ATGCCTAGGTACGGCCACGGTCAGATAGGTATACATTGCACTAGGCGTCGATATGTATACTAGCCATA

                                              AGCCGATGGGTAGATGGTC3'
                                              TCGGCTACCCATCTACCAG5'

S₁ → cloudy
5'TCTCTATGGACGATAGCAATATATGCAATCTTATCCCGATGACTACATCCATTCAGACTATGATCCA
3'AGAGATACCTGCTATCGTTATATACGTTAGAATAGGGCTACTGATGTAGGTAAGTCTGATACTAGGT

TGCCGCGTAACGATTATCCATGCCGATCGATAGCCAGATCCATGCCGATCCATGCGTAGCGATCCATG
ACGGCGCATTGCTAATAGGTACGGCTAGCTATCGGTCTAGGTACGGCTAGGTACGCATCGCTAGGTAC

CCGATCCATTACCGGCCGGATCCATGCCGATCCATGCTACCATAAGAATCCATGCCGATCCATGCCG3'
GGCTAGGTAATGGCCGGCCTAGGTACGGCTAGGTACGATGGTATTCTTAGGTACGGCTAGGTACGGC5'

C₁ (Consequence: visibility) → very high
5'CTACCATCAGATTCGGTGATGGATCCATGAGCATAGGGATCCATGCCGCTAACTTGTAATCCATGCC
3'GATGGTAGTCTAAGCCACTACCTAGGTACTCGTATCCCTAGGTACGGCGATTGAACATTAGGTACGG

                                        GATCCGTCTGATACGATCCATGCCGATCCATGCCG3'
                                        CTAGGCAGACTATGCTAGGTACGGCTAGGTACGGC5'

Other rules are similarly coded in the form of double stranded DNA sequences.

### 5.2. Statement of the problem:

The following values are given:

Dew point ($Td_{ob}$) : 25°C  
Dew point spread ($\Delta T_{ob}$) : -4.0°C  
Rate of change of dew point spread ($\Delta T'_{ob}$) : -2.0°C  
Wind speed ($W_{ob}$) : 7.0Knts/hr.  
Sky condition ($S_{ob}$) : 10.1%  
Possibility of visibility : ?

## 5.3. Wet lab algorithm

Step 1:
N number of expert rules are given, each of which has variable numbers of antecedent clauses and a single consequent clause. Finally, the observed data from the field are given in the fuzzified form for which the consequence has to be drawn.

Step 2:
The similarity between each of the antecedent clause of every rule and fuzzified observed data of the same discrete universe is measured. A threshold value is set for similarity measurement. The clauses having the degree of similarity greater than the threshold value are considered as dissimilar. In case all the antecedent clauses are similar to the fuzzified version of the observed data then the consequence of that particular rule is directly taken without any modification. The rules from the database are not considered for firing if none of the antecedent clause is similar to the fuzzified version of observed data of the respective domain. The rule whose at least one antecedent clause is similar to the observed data of the same discrete universe, is selected. Once a rule is selected, its highest degree of dissimilarity between the observed data and any of the antecedent clauses of the same universes of the selected rule is also recorded for modification of the membership value of the consequence of the selected rule as per the procedure of table 9. At this step essentially amplification, gel electrophoresis and modification of the membership values of the consequences of the selected rules are performed. At this stage, massive parallelism is adopted to exploit the merit of DNA computing.

Step 3:
From the modified consequences the final conclusion has to be derived according to the highest membership value.

## 5.4. Implementation of wet lab algorithm

Step 1 of the algorithm:

**a.** *At first we have to find the range of the domains in which the corresponding values lie.*

According the problem, Td (value 25°C) is under the primary fuzzy set 'very moist', as for this range (25≤Td≤30) 'very moist' has the highest membership value (i.e. 1.0).

According the problem, ΔT (value -4°C) is under the primary fuzzy set 'saturated', as for this range (-4≤ΔT<-2) 'saturated' has the highest membership value (i.e. 0.8).

According the problem, ΔT ' (value -2°C) is under the primary fuzzy set 'saturating', as for this range (-2≤ΔT<-1) 'saturating' has the highest membership value (i.e. 0.7).

According the problem, W (value 7Knts/hr.) is under the primary fuzzy set 'excellent', as for this range (5≤W<7.5) 'excellent' has the highest membership value (i.e. 0.9).

According the problem, S (value 10.1%) is under the primary fuzzy set 'clear', as for this range (10≤S<20) 'clear' has the highest membership value (i.e. 0.9).

**b.** *Encode the given values of antecedent parts by double stranded DNA sequence (i.e. fuzzy DNA).The data is used to generate the primed fuzzy set using triangular fuzzification. Triangular fuzzification ensures that the new primed fuzzy set which is generated, has a gradation in the membership values by filling adjacent partitions with a membership value 0.5 and partition in which it belongs with the membership value 1, thus, prohibiting the membership values to plummet to zero in an abrupt manner.*

By applying the method of triangular fuzzification (*Ray and Ghoshal, 2000*), we can state that, for the given values of antecedent parts, the sequences representing the corresponding ranges (table 2, 3, 4, 5 and 6) should have the membership value 1. The particular preceding and following two ranges have the membership values 0.9, 0.8, 0.7 and so on. All the short DNA oligonucleotides should have the oligonucleotides representing the corresponding membership values (table 1).

***Fuzzy DNA of Td = 25°C (Td$_{ob}$)***
5'CTGGACTGGATAATTTAATTGATCCGATCCATATTTTATTTTATCCTCAGCTCAGCATCCATCGAA
3'GACCTGACCTATTAAATTAACTAGGCTAGGTATAAAATAAAATAGGAGTCGAGTCGTAGGTAGCTT

TCGAATATCCATGCAATGTAATGTATCCATGCCGCCAGACCAGAATCCATGCCGATATCGTATCGTAT
AGCTTATAGGTACGTTACATTACATAGGTACGGCGGTCTGGTCTTAGGTACGGCTATAGCATAGCATA

CCATGCCGATCCTTAGATTAGAATCCATGCCGATCCATATTCAATTCAATCCATGCCGATCCATGCGC
GGTACGGCTAGGAATCTAATCTTAGGTACGGCTAGGTATAAGTTAAGTTAGGTACGGCTAGGTACGCG

                                                          CAAGCCAAATCCATGCCGATCCATGCCG3'
                                                          GTTCGGTTTAGGTACGGCTAGGTACGGC5'

***Fuzzy DNA of ΔT = -4.0°C (ΔT$_{ob}$)***
5'TTCGTTTCGTATCCATGCCGATCAAACCAAACATCCATGCCGATCCCGGAACGGAAATCCATGCCGA
3'AAGCAAAGCATAGGTACGGCTAGTTTGGTTTGTAGGTACGGCTAGGGCCTTGCCTTTAGGTACGGCT

TCCATATCCGATCCGATCCATGCCGATCCATGCGAATGGAATGATCCATGCCGATCCATGCCGGTAGC
AGGTATAGGCTAGGCTAGGTACGGCTAGGTACGCTTACCTTACTAGGTACGGCTAGGTACGGCCATCG

```
GTAGCATCCATGCCGATCCATGCATCCCATCCCATCCATGCCGATCCATTAGGATAGGAATCCATGCC
CATCGTAGGTACGGCTAGGTACGTAGGGTAGGGTAGGTACGGCTAGGTAATCCTATCCTTAGGTACGG

GATCCCTAAGCTAAGATCCATGCCGATAGGAAAGGAAATCCATGCCGTAGCTTAGCTATCCATGCGCG
CTAGGGATTCGATTCTAGGTACGGCTATCCTTTCCTTAGGTACGGCATCGAATCGATAGGTACGCGC

CGGCGCGATCCAT3'
GCCGCGCTAGGTA5'
```

*Fuzzy DNA of ΔT ' = -2.0°C (ΔT 'ₒᵦ)*
```
5'GTAACGTAACATCCATGCCGATCCAAATAAAATAATCCATGCCGATCCATTGCGATGATTATCCATG
3'CATTGCATTGTAGGTACGGCTAGGTTTATTTTATTAGGTACGGCTAGGTAACGCTACTAATAGGTAC

CCGATCCATGCCGGAATTAACATCCATGCCGATCCATGCCGGTAATCTATGATCCATGCCGGTTAACG
GGCTAGGTACGGCCTTAATTGTAGGTACGGCTAGGTACGGCCATTAGATACTAGGTACGGCCAATTGC

            GTAAGTAGCGTAGTACCGTAACTAGCGTAGCTACAGTACCGATTACGCAGGAA3'
            CATTCATCGCATCATGGCATTGATCGCATCGATGTCATGGCTAATGCGTCCTT5'
```

*Fuzzy DNA of W = 7.0Knts/hr. (Wₒᵦ)*
```
5'GCTATGGCTGATCCATGCCGATATACCGATAGATCCATGCCGATCCTAACCCGTGGATCCATGCCGA
3'CGATACCGACTAGGTACGGCTATATGGCTATCTAGGTACGGCTAGGATTGGGCACCTAGGTACGGCT

TCCATCAGTAAGTCTATCCATGCCGATCCATGCATTCCGATGCATCCATGCCGATCCATGCCGTACAG
AGGTAGTCATTCAGATAGGTACGGCTAGGTACGTAAGGCTACGTAGGTACGGCTAGGTACGGCATGTC

TACGGATCCATGCCGATCCATGCGTGCCAGTCTATCCATGCCGATCCATATGTAACGTGATCCATGCC
ATGCCTAGGTACGGCTAGGTACGCACGGTCAGATAGGTACGGCTAGGTATACATTGCACTAGGTACGG

GATCCGCAGCTATACATCCATGCCGATATGATCGGTAATCCATGCCGTAGCCGATGGATCCATGCGTA
CTAGGCGTCGATATGTAGGTACGGCTATACTAGCCATTAGGTACGGCATCGGCTACCTAGGTACGCAT

                                                        GATGGTCATCCAT3'
                                                        CTACCAGTAGGTA5'
```

*Fuzzy DNA of S = 10.1% (Sₒᵦ)*
```
5'TCTCTATGGAATCCATGCCGATCCATGCCGATAGCAATATCCATGCCGATCCATGCCGATGCAATCT
3'AGAGATACCTTAGGTACGGCTAGGTACGGCTATCGTTATAGGTACGGCTAGGTACGGCTACGTTAGA

TATCCATGCCGATCCATGCCGATGACTACATCCATGCCGATCCATTCAGACTATGATCCATGCCGATC
ATAGGTACGGCTAGGTACGGCTACTGATGTAGGTACGGCTAGGTAAGTCTGATACTAGGTACGGCTAG

CCGTAACGATTATCCATGCCGATCGATAGCCAGATCCATGCCGATGCGTAGCGATCCATGCTACCGGC
GGCATTGCTAATAGGTACGGCATGCTATCGGTCTAGGTACGGCTACGCATCGCTAGGTACGATGGCCG

CGGATCCATTACCATAAGAATCC3'
GCCTAGGTAATGGTATTCTTAGG5'
```

Step 2 of the algorithm:

**a.** *Each of the sequences of all the domains is amplified by PCR using specific primers.*

**Primer Used:**

*For Td Domain:*
  i. The complementary to the short DNA sequence representing the range having the highest membership value for the particular primary fuzzy set which is desired to amplify.
  ii. 5'-CTGGACTGGA-3' (i.e. the short DNA sequence representing the particular range having membership value 1.0 for the primary fuzzy set 'dry'.

*For ΔT Domain:*
  i. The complementary to the short DNA sequence representing the range having the highest membership value for the particular primary fuzzy set which is desired to amplify.
  ii. 5'-TTCGTTTCGT-3' (i.e. the short DNA sequence representing the particular range having membership value 1.0 for the primary fuzzy set 'very saturated').

*For ΔT ' Domain:*
  i. The complementary to the short DNA sequence representing the range having the highest membership value for the particular primary fuzzy set which is desired to amplify.
  ii. 5'-GTAACGTAAC-3' (i.e. the short DNA sequence representing the particular range having membership value 1.0 for the primary fuzzy set 'saturating').

*For W Domain:*
  i. The complementary to the short DNA sequence representing the range having the highest membership value for the particular primary fuzzy set which is desired to amplify.
  ii. 5'-GCTATGGCTG-3' (i.e. the short DNA sequence representing the particular range having membership value 1.0 for the primary fuzzy set 'too light').

*For S Domain:*
  i. The complementary to the short DNA sequence representing the range having the highest membership value for the particular primary fuzzy set which is desired to amplify.
  ii. 5'-TCTCTATGGA-3' (i.e. the short DNA sequence representing the particular range having membership value 1.0 for the primary fuzzy set 'clear').

Rule 1:

For $Td_1 \rightarrow dry$ primers used: 5'CTGGACTGGA3' and 3'GACCTGACCT5'

Amplification of 5'-3' strand:
5'***CTGGACTGGA***ATCCATGCCGATCCATGCCGTAATTTAATTATCCATGCCGATCCATGCGATCCGATC
3'***GACCTGACCT5'***

CATCCATGCCGATCCATATTTTATTTTATCCATGCCGATCCTCAGCTCAGCATCCATGCCGATCGAA

TCGAATATCCATGCCGAATGTAATGTATCCATCCAGACCAGAATCCATCGTATCGTATTTAGATTAGAA

TTCAATTCAGCCAAGCCAA3'

Amplification of 3'-5' strand:
3'***GACCTGACCT***TAGGTACGGCTAGGTACGGCATTAAATTAATAGGTACGGCTAGGTACGCTAGGCTAG
5'***CTGGACTGGA3'***

GTAGGTACGGCTAGGTATAAAATAAAATAGGTACGGCTAGGAGTCGAGTCGTAGGTACGGCTAGCTT

AGCTTATAGGTACGGCTTACATTACATAGGTAGGTCTGGTCTTAGGTAGCATAGCATAAATCTAATCTT

                                                              AAGTTAAGTCGGTTCGGTT5'

**The length of the amplified sequence is 10 bp.**
*5'CTGGACTGGA3'*
*3'GACCTGACCT5'*

For *ΔT₁* → *very saturated* primers used: 5'TTCGTTTCGT3' and 3'AAGCAAAGCA5'

Amplification of 5'-3' strand:
*5'TTCGTTTCGT*ATCCATGCCGATCCATGCCGCAAACCAAACATCCATGCCGATCCATGCCGGAACGG
*3'AAGCAAAGCA5'*

AAATCCATGCCGATCCATATCCGATCCGATCCATGCCGATCCGAATGGAATGATCCATGCCGATGTAG

CGTAGCATCCATGCCGATCCCATCCCATCCATTAGGATAGGAATCCCTAAGCTAAGATAGGAAAGGAA

                                                            TAGCTTAGCTGCGCGGCGCG3'

Amplification of 3'-5' strand:
*3'AAGCAAAGCA*TAGGTACGGCTAGGTACGGCGTTTGGTTTGTAGGTACGGCTAGGTACGGCCTTGCC
*5'TTCGTTTCGT3'*

TTTAGGTACGGCTAGGTATAGGCTAGGCTAGGTACGGCTAGGCTTACCTTACTAGGTACGGCTACATC

GCATCGTAGGTACGGCTAGGGTAGGGTAGGTAATCCTATCCTTAGGGATTCGATTCTATCCTTTCCTT

                                                            ATCGAATCGACGCGCCGCGC5'

**The length of the amplified sequence is 10 bp.**
*5'TTCGTTTCGT3'*
*3'AAGCAAAGCA5'*

For *ΔT₁'* → *drying* primers used: 5'GTAACGTAAC3' and 3'TGCGTCCTT5'

Amplification of 5'-3' strand:
*5'GTAACGTAACAAATAAAATATGCGATGATT*ATCGGAATTAAC*ATCCGTAATCTATGATCCATGTTAA*
........................................................................................................................................................

*CGGTA*ATCCATGCCGAGTAGCGTAG*ATCCATGCCGATT*ACCGTAACT*ATCCATGCCGATCCAGCGTAG*
........................................................................................................................................................

*CTA*ATCCATGCCGATCCAT*CAGTACCGATT*ATCCATGCCGATCCATGCACGCAGGAAATCCATGCCGA
........................................................................................←*3'TGCGTCCTT5'*

                                                                   TCCATGCCG3'

Amplification of 3'-5' strand:
*3'CATTGCATTGTTTATTTTATACGCTACTAATAGCCTTAATTGTAGGCATTAGATACTAGGTACAATT*
*5'GTAACGTAAC3'→*..................................................................................................

*GCCATTAGGTACGGCTCATCGCATCTAGGTACGGCTAATGGCATTGATAGGTACGGCTAGGTCGCATC*
...........................................................................................................................

*GATTAGGTACGGCTAGGTAGTCATGGCTAATAGGTACGGCTAGGTACGTGCGTCCTT*TAGGTACGGCT
...............................................................................................................

AGGTACGGC5'

**The length of the amplified sequence is 192 bp.**
*5'GTAACGTAACAAATAAAATATGCGATGATTATCGGAATTAACATCCGTAATCTATGATCCATGTTAA*
*3'CATTGCATTGTTTATTTTATACGCTACTAATAGCCTTAATTGTAGGCATTAGATACTAGGTACAATT*

*CGGTAATCCATGCCGAGTAGCGTAGATCCATGCCGATTACCGTAACTATCCATGCCGATCCAGCGTAG*
*GCCATTAGGTACGGCTCATCGCATCTAGGTACGGCTAATGGCATTGATAGGTACGGCTAGGTCGCATC*

*CTAATCCATGCCGATCCATCAGTACCGATTATCCATGCCGATCCATGCACGCAGGAA3'*
*GATTAGGTACGGCTAGGTAGTCATGGCTAATAGGTACGGCTAGGTACGTGCGTCCTT5'*

For $W_1 \rightarrow$ *too light* primers used: 5'GCTATGGCTG3' and 3'CGATACCGAC5'

Amplification of 5'-3' strand:
*5'GCTATGGCTG*ATCCATGCCGATCCATGCCGATACCGATAGATCCATGCCGATCCATGCTAACCCGTG
*3'CGATACCGAC5'*

GATCCATGCCGATCCATCAGTAAGTCTATCCATGCCGATCCATTCCGATGCATCCATGCCGATTACAG

TACGGATCCATGCCGGTGCCAGTCTATCCATATGTAACGTGATCCGCAGCTATACATATGATCGGTAT

AGCCGATGGGTAGATGGTC3'

Amplification of 3'-5' strand:
*3'CGATACCGAC*TAGGTACGGCTAGGTACGGCTATGGCTATCTAGGTACGGCTAGGTACGATTGGGCAC
*5'GCTATGGCTG3'*

CTAGGTACGGCTAGGTAGTCATTCAGATAGGTACGGCTAGGTAAGGCTACGTAGGTACGGCTAATGTC

ATGCCTAGGTACGGCCACGGTCAGATAGGTATACATTGCACTAGGCGTCGATATGTATACTAGCCATA

TCGGCTACCCATCTACCAG5'

**The length of the amplified sequence is 10 bp.**
*5'GCTATGGCTG3'*
*3'CGATACCGAC5'*

For $S_1 \rightarrow$ *cloudy* primers used: 5'TCTCTATGGA3' and 3'ATGGTATTCT5'

Amplification of 5'-3' strand:
*5'TCTCTATGGACGATAGCAATATATGCAATCTTATCCCGATGACTACATCCATTCAGACTATGATCCA*

..............................................................................................................................................................
*TGCCGCGTAACGATTATCCATGCCGATCGATAGCCAGATCCATGCCGATCCATGCGTAGCGATCCATG*
..............................................................................................................................................................

*CCGATCCATTACCGGCCGGATCCATGCCGATCCATGCTACCATAAGA*ATCCATGCCGATCCATGCCG3'
..........................................................................................←*3'ATGGTATTCT5'*

Amplification of 3'-5' strand:
*3'AGAGATACCTGCTATCGTTATATACGTTAGAATAGGGCTACTGATGTAGGTAAGTCTGATACTAGGT*
*5'TCTCTATGGA3'*→ ..............................................................................................................................................

*ACGGCGCATTGCTAATAGGTACGGCTAGCTATCGGTCTAGGTACGGCTAGGTACGCATCGCTAGGTAC*
..............................................................................................................................................................

*GGCTAGGTAATGGCCGGCCTAGGTACGGCTAGGTACGATGGTATTCT*TAGGTACGGCTAGGTACGGC5'
..............................................................................................................................................................

**The length of the amplified sequence is 182 bp.**
*5'TCTCTATGGACGATAGCAATATATGCAATCTTATCCCGATGACTACATCCATTCAGACTATGATCCA*
*3'AGAGATACCTGCTATCGTTATATACGTTAGAATAGGGCTACTGATGTAGGTAAGTCTGATACTAGGT*

*TGCCGCGTAACGATTATCCATGCCGATCGATAGCCAGATCCATGCCGATCCATGCGTAGCGATCCATG*
*ACGGCGCATTGCTAATAGGTACGGCTAGCTATCGGTCTAGGTACGGCTAGGTACGCATCGCTAGGTAC*

*CCGATCCATTACCGGCCGGATCCATGCCGATCCATGCTACCATAAGA*
*GGCTAGGTAATGGCCGGCCTAGGTACGGCTAGGTACGATGGTATTCT5'*

Likewise,

*Rule 2:*

For *Td₂ → dry*:
**The length of the amplified sequence is 10 bp.**
*5'CTGGACTGGA3'*
*3'GACCTGACCT5'*

For *ΔT₂ → saturated*:
**The length of the amplified sequence is 146 bp.**
*5'TTCGTTTCGTATCCATCAAACCAAACATCCATGCCGCGGAACGGAAATCCATGCCGATATCCGATCC*
*3'AAGCAAAGCATAGGTAGTTTGGTTTGTAGGTACGGCGCCTTGCCTTTAGGTACGGCTATAGGCTAGG*

*GATCCATGCCGATCCGAATGGAATGATCCATGCCGATCCATGTAGCGTAGCATCCATGCCGATCCATG*
*CTAGGTACGGCTAGGCTTACCTTACTAGGTACGGCTAGGTACATCGCATCGTAGGTACGGCTAGGTAC*

*CATCCCATCCC3'*
*GTAGGGTAGGG5'*

For *ΔT₂' → saturating*:
**The length of the amplified sequence is 10 bp.**
*5'GTAACGTAAC3'*
*3'CATTGCATTG5'*

For *W₂ → excellent*:

**The length of the amplified sequence is 130 bp.**
*5'GCTATGGCTGATCCATGCCGATACCGATAGATCCATGCCGATTAACCCGTGGATCCATGCCGATCCC*
*3'CGATACCGACTAGGTACGGCTATGGCTATCTAGGTACGGCTAATTGGGCACCTAGGTACGGCTAGGG*

*AGTAAGTCTATCCATGCCGATCCATATTCCGATGCATCCATGCCGATCCATGCTACAGTACGG3'*
*TCATTCAGATAGGTACGGCTAGGTATAAGGCTACGTAGGTACGGCTAGGTACGATGTCATGCC5'*

For $S_2 \to$ *clear*:
**The length of the amplified sequence is 10 bp.**
*5'TCTCTATGGA3'*
*3'AGAGATACCT5'*

*Rule 3:*

For $Td_3 \to$ *moderate*:
**The length of the amplified sequence is 130 bp.**
*5'CTGGACTGGAATCCATGCCGTAATTTAATTATCCATGCCGATGATCCGATCCATCCATGCCGATCC*
*3'GACCTGACCTTAGGTACGGCATTAAATTAATAGGTACGGCTACTAGGCTAGGTAGGTACGGCTAGG*

*ATTTTATTTTATCCATGCCGATCCATTCAGCTCAGCATCCATGCCGATCCATGCCGAATCGAAT3'*
*TAAAATAAAATAGGTACGGCTAGGAGTCGAGTCGTAGGTACGGCTAGGTACGGCTTAGCTTA5'*

For $\Delta T_3 \to$ *unsaturated*:
**The length of the amplified sequence is 202 bp.**
*5'TTCGTTTCGTCAAACCAAACCGGAACGGAAATCCGATCCGATGAATGGAATGATCCGTAGCGTAGCA*
*3'AAGCAAAGCAGTTTGGTTTGGCCTTGCCTTTAGGCTAGGCTACTTACCTTACTAGGCATCGCATCGT*

*TCCATATCCCATCCCATCCATGCCGTAGGATAGGAATCCATGCCGATCTAAGCTAAGATCCATGCCGA*
*AGGTATAGGGTAGGGTAGGTACGGCATCCTATCCTTAGGTACGGCTAGATTCGATTCTAGGTACGGCT*

*TCCAGGAAAGGAAATCCATGCCGATCCATTAGCTTAGCTATCCATGCCGATCCATGCGCGCGGCGCG5'*
*AGGTCCTTTCCTTTAGGTACGGCTAGGTAATCGAATCGATAGGTACGGCTAGGTACGCGCGCCGCGC3'*

For $\Delta T_3' \to$ *drying*:
**The length of the amplified sequence is 192 bp.**
*5'GTAACGTAACAAATAAAATATGCGATGATTATCGGAATTAACATCCGTAATCTATGATCCATGTTAA*
*3'CATTGCATTGTTTATTTTATACGCTACTAATAGCCTTAATTGTAGGCATTAGATACTAGGTACAATT*

*CGGTAATCCATGCCGAGTAGCGTAGATCCATGCCGATTACCGTAACTATCCATGCCGATCCAGCGTAG*
*GCCATTAGGTACGGCTCATCGCATCTAGGTACGGCTAATGGCATTGATAGGTACGGCTAGGTCGCATC*

*CTAATCCATGCCGATCCATCAGTACCGATTATCCATGCCGATCCATGCACGCAGGAA3'*
*GATTAGGTACGGCTAGGTAGTCATGGCTAATAGGTACGGCTAGGTACGTGCGTCCTT5'*

For $W_3 \to$ *too strong*:
**The length of the amplified sequence is 202 bp.**
*5'GCTATGGCTGATACCGATAGTAACCCGTGGCAGTAAGTCTATATTCCGATGCATCCTACAGTACGGA*
*3'CGATACCGACTATGGCTATCATTGGGCACCGTCATTCAGATATAAGGCTACGTAGGATGTCATGCCT*

*TCCATGTGCCAGTCTATCCATGCCGATGTAACGTGATCCATGCCGATGCAGCTATACATCCATGCCGA*
*AGGTACACGGTCAGATAGGTACGGCTACATTGCACTAGGTACGGCTACGTCGATATGTAGGTACGGCT*

*TCCATGATCGGTAATCCATGCCGATCCATTAGCCGATGGATCCATGCCGATCCATGCGTAGATGGTC3'*
*AGGTACTAGCCATTAGGTACGGCTAGGTAATCGGCTACCTAGGTACGGCTAGGTACGCATCTACCAG5'*

For *S₃ → partially cloudy*:
**The length of the amplified sequence is 110 bp.**
*5'TCTCTATGGAATCCATGCCGATCGATAGCAATATCCATGCCGATCCATGCAATCTTATCCATGCCGA*
*3'AGAGATACCTTAGGTACGGCTAGCTATCGTTATAGGTACGGCTAGGTACGTTAGAATAGGTACGGCT*

*TCCATCGATGACTACATCCATGCCGATCCATGCTCAGACTATG3'*
*AGGTAGCTACTGATGTAGGTACGGCTAGGTACGAGTCTGATAC5'*

*Rule 4:*

For *Td₄ → moist:*
**The length of the amplified sequence is 158 bp.**
*5'CTGGACTGGAATTAATTTAATTATCCATGATCCGATCCATCCATGCCGATTTTATTTTATCCATGCC*
*3'GACCTGACCTTAATTAAATTAATAGGTACTAGGCTAGGTAGGTACGGCTAAAATAAAATAGGTACGG*

*GATTCAGCTCAGCATCCATGCCGATCCCGAATCGAATATCCATGCCGATCCATAATGTAATGTATCCA*
*CTAAGTCGAGTCGTAGGTACGGCTAGGGCTTAGCTTATAGGTACGGCTAGGTATTACATTACATAGGT*

*TGCCGATCCATGCCCAGACCAGA3'*
*ACGGCTAGGTACGGGTCTGGTCT5'*

For *ΔT₄ → saturated*:
**The length of the amplified sequence is 146 bp.**
*5'TTCGTTTCGTATCCATCAAACCAAACATCCATGCCGCGGAACGGAAATCCATGCCGATATCCGATCC*
*3'AAGCAAAGCATAGGTAGTTTGGTTTGTAGGTACGGCGCCTTGCCTTTAGGTACGGCTATAGGCTAGG*

*GATCCATGCCGATCCGAATGGAATGATCCATGCCGATCCATGTAGCGTAGCATCCATGCCGATCCATG*
*CTAGGTACGGCTAGGCTTACCTTACTAGGTACGGCTAGGTACATCGCATCGTAGGTACGGCTAGGTAC*

*CATCCCATCCC3'*
*GTAGGGTAGGG5'*

For *ΔT₄' → drying*:
**The length of the amplified sequence is 192 bp.**
*5'GTAACGTAACAAATAAAATATGCGATGATTATCGGAATTAACATCCGTAATCTATGATCCATGTTAA*
*3'CATTGCATTGTTTATTTTATACGCTACTAATAGCCTTAATTGTAGGCATTAGATACTAGGTACAATT*

*CGGTAATCCATGCCGAGTAGCGTAGATCCATGCCGATTACCGTAACTATCCATGCCGATCCAGCGTAG*
*GCCATTAGGTACGGCTCATCGCATCTAGGTACGGCTAATGGCATTGATAGGTACGGCTAGGTCGCATC*

*CTAATCCATGCCGATCCATCAGTACCGATTATCCATGCCGATCCATGCACGCAGGAA3'*
*GATTAGGTACGGCTAGGTAGTCATGGCTAATAGGTACGGCTAGGTACGTGCGTCCTT5'*

For *W₄ → excellent*:
**The length of the amplified sequence is 130 bp.**
*5'GCTATGGCTGATCCATGCCGATACCGATAGATCCATGCCGATTAACCCGTGGATCCATGCCGATCCC*
*3'CGATACCGACTAGGTACGGCTATGGCTATCTAGGTACGGCTAATTGGGCACCTAGGTACGGCTAGGG*

*AGTAAGTCTATCCATGCCGATCCATATTCCGATGCATCCATGCCGATCCATGCTACAGTACGG3'*
*TCATTCAGATAGGTACGGCTAGGTATAAGGCTACGTAGGTACGGCTAGGTACGATGTCATGCC5'*

For *S₄ → clear*:
**The length of the amplified sequence is 10 bp.**

| |
|---|
| *5'TCTCTATGGA3'*  *3'AGAGATACCT5'* |

*Rule 5:*

For *Td$_5$ → moist*:
**The length of the amplified sequence is 158 bp.**
*5'CTGGACTGGAATTAATTTAATTATCCATGATCCGATCCATCCATGCCGATTTTATTTTATCCATGCC*
*3'GACCTGACCTTAATTAAATTAATAGGTACTAGGCTAGGTAGGTACGGCTAAAATAAAATAGGTACGG*

*GATTCAGCTCAGCATCCATGCCGATCCCGAATCGAATATCCATGCCGATCCATAATGTAATGTATCCA*
*CTAAGTCGAGTCGTAGGTACGGCTAGGGCTTAGCTTATAGGTACGGCTAGGTATTACATTACATAGGT*

*TGCCGATCCATGCCCAGACCAGA3'*
*ACGGCTAGGTACGGGTCTGGTCT5'*

For *ΔT$_5$ → unsaturated*:
**The length of the amplified sequence is 202 bp.**
*5'TTCGTTTCGTCAAACCAAACCGGAACGGAAATCCGATCCGATGAATGGAATGATCCGTAGCGTAGCA*
*3'AAGCAAAGCAGTTTGGTTTGGCCTTGCCTTTAGGCTAGGCTACTTACCTTACTAGGCATCGCATCGT*

*TCCATATCCCATCCCATCCATGCCGTAGGATAGGAATCCATGCCGATCTAAGCTAAGATCCATGCCGA*
*AGGTATAGGGTAGGGTAGGTACGGCATCCTATCCTTAGGTACGGCTAGATTCGATTCTAGGTACGGCT*

*TCCAGGAAAGGAAATCCATGCCGATCCATTAGCTTAGCTATCCATGCCGATCCATGCGCGCGGCGCG5'*
*AGGTCCTTTCCTTTAGGTACGGCTAGGTAATCGAATCGATAGGTACGGCTAGGTACGCGCGCCGCGC3'*

For *ΔT$_5$' → saturating*:
**The length of the amplified sequence is 10 bp.**
*5'GTAACGTAAC3'*
*3'CATTGCATTG5'*

For *W$_5$ → too strong*:
**The length of the amplified sequence is 202 bp.**
*5'GCTATGGCTGATACCGATAGTAACCCGTGGCAGTAAGTCTATATTCCGATGCATCCTACAGTACGGA*
*3'CGATACCGACTATGGCTATCATTGGGCACCGTCATTCAGATATAAGGCTACGTAGGATGTCATGCCT*

*TCCATGTGCCAGTCTATCCATGCCGATGTAACGTGATCCATGCCGATGCAGCTATACATCCATGCCGA*
*AGGTACACGGTCAGATAGGTACGGCTACATTGCACTAGGTACGGCTACGTCGATATGTAGGTACGGCT*

*TCCATGATCGGTAATCCATGCCGATCCATTAGCCGATGGATCCATGCCGATCCATGCGTAGATGGTC3'*
*AGGTACTAGCCATTAGGTACGGCTAGGTAATCGGCTACCTAGGTACGGCTAGGTACGCATCTACCAG5'*

For *S$_5$ → cloudy*:
**The length of the amplified sequence is 182 bp.**
*5'TCTCTATGGACGATAGCAATATATGCAATCTTATCCCGATGACTACATCCATTCAGACTATGATCCA*
*3'AGAGATACCTGCTATCGTTATATACGTTAGAATAGGGCTACTGATGTAGGTAAGTCTGATACTAGGT*

*TGCCGCGTAACGATTATCCATGCCGATCGATAGCCAGATCCATGCCGATCCATGCGTAGCGATCCATG*
*ACGGCGCATTGCTAATAGGTACGGCTAGCTATCGGTCTAGGTACGGCTAGGTACGCATCGCTAGGTAC*

*CCGATCCATTACCGGCCGGATCCATGCCGATCCATGCTACCATAAGA*
*GGCTAGGTAATGGCCGGCCTAGGTACGGCTAGGTACGATGGTATTCT5'*

*Rule 6:*

For *Td₆ → very moist*:
**The length of the amplified sequence is 202 bp.**
*5'CTGGACTGGATAATTTAATTGATCCGATCCATTTTATTTTATTCAGCTCAGCATCCCGAATCGAATA*
*3'GACCTGACCTATTAAATTAACTAGGCTAGGTAAAATAAAATAAGTCGAGTCGTAGGGCTTAGCTTAT*

*TCCATAATGTAATGTATCCATGCCGCCAGACCAGAATCCATGCCGATATCGTATCGTATCCATGCCGA*
*AGGTATTACATTACATAGGTACGGCGGTCTGGTCTTAGGTACGGCTATAGCATAGCATAGGTACGGCT*

*TCCTTAGATTAGAATCCATGCCGATCCATATTCAATTCAATCCATGCCGATCCATGCGCCAAGCCAA5'*
*AGGAATCTAATCTTAGGTACGGCTAGGTATAAGTTAAGTTAGGTACGGCTAGGTACGCGGTTCGGTT3'*

For *ΔT₆ → very saturated*:
**The length of the amplified sequence is 10 bp.**
*5'TTCGTTTCGT3'*
*3'AAGCAAAGCA5'*

For *ΔT₆' → saturating*:
**The length of the amplified sequence is 10 bp.**
*5'GTAACGTAAC3'*
*3'CATTGCATTG5'*

For *W₆ → too light*:
**The length of the amplified sequence is 10 bp.**
*5'GCTATGGCTG3'*
*3'CGATACCGAC5'*

For *S₆ → clear*:
**The length of the amplified sequence is 10 bp.**
*5'TCTCTATGGA3'*
*3'AGAGATACCT5'*

*Rule 7:*

For *Td₇ → very moist*:
**The length of the amplified sequence is 202 bp.**
*5'CTGGACTGGATAATTTAATTGATCCGATCCATTTTATTTTATTCAGCTCAGCATCCCGAATCGAATA*
*3'GACCTGACCTATTAAATTAACTAGGCTAGGTAAAATAAAATAAGTCGAGTCGTAGGGCTTAGCTTAT*

*TCCATAATGTAATGTATCCATGCCGCCAGACCAGAATCCATGCCGATATCGTATCGTATCCATGCCGA*
*AGGTATTACATTACATAGGTACGGCGGTCTGGTCTTAGGTACGGCTATAGCATAGCATAGGTACGGCT*

*TCCTTAGATTAGAATCCATGCCGATCCATATTCAATTCAATCCATGCCGATCCATGCGCCAAGCCAA5'*
*AGGAATCTAATCTTAGGTACGGCTAGGTATAAGTTAAGTTAGGTACGGCTAGGTACGCGGTTCGGTT3'*

For *ΔT₇ → saturated*:
**The length of the amplified sequence is 146 bp.**
*5'TTCGTTTCGTATCCATCAAACCAAACATCCATGCCGCGGAACGGAAATCCATGCCGA*
*TATCCGATCC*
*3'AAGCAAAGCATAGGTAGTTTGGTTTGTAGGTACGGCGCCTTGCCTTTAGGTACGGCTATAGGCTAGG*

*GATCCATGCCGATCCGAATGGAATGATCCATGCCGATCCATGTAGCGTAGCATCCATGCCGATCCATG*
*CTAGGTACGGCTAGGCTTACCTTACTAGGTACGGCTAGGTACATCGCATCGTAGGTACGGCTAGGTAC*

*CATCCCATCCC3'*
*GTAGGGTAGGG5'*

For $\Delta T_7' \rightarrow drying$:
**The length of the amplified sequence is 192 bp.**
*5'GTAACGTAACAAATAAAATATGCGATGATTATCGGAATTAACATCCGTAATCTATGATCCATGTTAA*
*3'CATTGCATTGTTTATTTTATACGCTACTAATAGCCTTAATTGTAGGCATTAGATACTAGGTACAATT*

*CGGTAATCCATGCCGAGTAGCGTAGATCCATGCCGATTACCGTAACTATCCATGCCGATCCAGCGTAG*
*GCCATTAGGTACGGCTCATCGCATCTAGGTACGGCTAATGGCATTGATAGGTACGGCTAGGTCGCATC*

*CTAATCCATGCCGATCCATCAGTACCGATTATCCATGCCGATCCATGCACGCAGGAA3'*
*GATTAGGTACGGCTAGGTAGTCATGGCTAATAGGTACGGCTAGGTACGTGCGTCCTT5'*

For $W_7 \rightarrow too\ strong$:
**The length of the amplified sequence is 202 bp.**
*5'GCTATGGCTGATACCGATAGTAACCCGTGGCAGTAAGTCTATATTCCGATGCATCCTACAGTACGGA*
*3'CGATACCGACTATGGCTATCATTGGGCACCGTCATTCAGATATAAGGCTACGTAGGATGTCATGCCT*

*TCCATGTGCCAGTCTATCCATGCCGATGTAACGTGATCCATGCCGATGCAGCTATACATCCATGCCGA*
*AGGTACACGGTCAGATAGGTACGGCTACATTGCACTAGGTACGGCTACGTCGATATGTAGGTACGGCT*

*TCCATGATCGGTAATCCATGCCGATCCATTAGCCGATGGATCCATGCCGATCCATGCGTAGATGGTC3'*
*AGGTACTAGCCATTAGGTACGGCTAGGTAATCGGCTACCTAGGTACGGCTAGGTACGCATCTACCAG5'*

For $S_7 \rightarrow partially\ cloudy$:
**The length of the amplified sequence is 110 bp.**
*5'TCTCTATGGAATCCATGCCGATCGATAGCAATATCCATGCCGATCCATGCAATCTTATCCATGCCGA*
*3'AGAGATACCTTAGGTACGGCTAGCTATCGTTATAGGTACGGCTAGGTACGTTAGAATAGGTACGGCT*

*TCCATCGATGACTACATCCATGCCGATCCATGCTCAGACTATG3'*
*AGGTAGCTACTGATGTAGGTACGGCTAGGTACGAGTCTGATAC5'*

*Rule 8:*

For $Td_8 \rightarrow moderate$:
**The length of the amplified sequence is 130 bp.**
*5'CTGGACTGGAATCCATGCCGTAATTTAATTATCCATGCCGATGATCCGATCCATCCATGCCGATCC*
*3'GACCTGACCTTAGGTACGGCATTAAATTAATAGGTACGGCTACTAGGCTAGGTAGGTACGGCTAGG*

*ATTTTATTTTATCCATGCCGATCCATTCAGCTCAGCATCCATGCCGATCCATGCCGAATCGAAT3'*
*TAAAATAAAATAGGTACGGCTAGGAGTCGAGTCGTAGGTACGGCTAGGTACGGCTTAGCTTA5'*

For $\Delta T_8 \rightarrow saturated$:
**The length of the amplified sequence is 146 bp.**
*5'TTCGTTTCGTATCCATCAAACCAAACATCCATGCCGCGGAACGGAAATCCATGCCGATATCCGATCC*
*3'AAGCAAAGCATAGGTAGTTTGGTTTGTAGGTACGGCGCCTTGCCTTTAGGTACGGCTATAGGCTAGG*

*GATCCATGCCGATCCGAATGGAATGATCCATGCCGATCCATGTAGCGTAGCATCCATGCCGATCCATG*
*CTAGGTACGGCTAGGCTTACCTTACTAGGTACGGCTAGGTACATCGCATCGTAGGTACGGCTAGGTAC*

*CATCCCATCCC3'*
*GTAGGGTAGGG5'*

For *ΔT₈′* → *saturating*:
**The length of the amplified sequence is 10 bp.**
*5'GTAACGTAAC3'*
*3'CATTGCATTG5'*

For *W₈* → *excellent*:
**The length of the amplified sequence is 130 bp.**
*5'GCTATGGCTGATCCATGCCGATACCGATAGATCCATGCCGATTAACCCGTGGATCCATGCCGATCCC*
*3'CGATACCGACTAGGTACGGCTATGGCTATCTAGGTACGGCTAATTGGGCACCTAGGTACGGCTAGGG*

*AGTAAGTCTATCCATGCCGATCCATATTCCGATGCATCCATGCCGATCCATGCTACAGTACGG3'*
*TCATTCAGATAGGTACGGCTAGGTATAAGGCTACGTAGGTACGGCTAGGTACGATGTCATGCC5'*

For *S₈* → *cloudy*:
**The length of the amplified sequence is 182 bp.**
*5'TCTCTATGGACGATAGCAATATATGCAATCTTATCCCGATGACTACATCCATTCAGACTATGATCCA*
*3'AGAGATACCTGCTATCGTTATATACGTTAGAATAGGGCTACTGATGTAGGTAAGTCTGATACTAGGT*

*TGCCGCGTAACGATTATCCATGCCGATCGATAGCCAGATCCATGCCGATCCATGCGTAGCGATCCATG*
*ACGGCGCATTGCTAATAGGTACGGCTAGCTATCGGTCTAGGTACGGCTAGGTACGCATCGCTAGGTAC*

*CCGATCCATTACCGGCCGGATCCATGCCGATCCATGCTACCATAAGA*
*GGCTAGGTAATGGCCGGCCTAGGTACGGCTAGGTACGATGGTATTCT5'*

*Rule 9:*

For *Td₉* → *moderate*:
**The length of the amplified sequence is 130 bp.**
*5'CTGGACTGGAATCCATGCCGTAATTTAATTATCCATGCCGATGATCCGATCCATCCATGCCGATCC*
*3'GACCTGACCTTAGGTACGGCATTAAATTAATAGGTACGGCTACTAGGCTAGGTAGGTACGGCTAGG*

*ATTTTATTTTATCCATGCCGATCCATTCAGCTCAGCATCCATGCCGATCCATGCCGAATCGAAT3'*
*TAAAATAAAATAGGTACGGCTAGGAGTCGAGTCGTAGGTACGGCTAGGTACGGCTTAGCTTA5'*

For *ΔT₉* → *very saturated*:
**The length of the amplified sequence is 10 bp.**
*5'TTCGTTTCGT3'*
*3'AAGCAAAGCA5'*

For *ΔT₉′* → *drying*:
**The length of the amplified sequence is 192 bp.**
*5'GTAACGTAACAAATAAAATATGCGATGATTATCGGAATTAACATCCGTAATCTATGATCCATGTTAA*
*3'CATTGCATTGTTTATTTTATACGCTACTAATAGCCTTAATTGTAGGCATTAGATACTAGGTACAATT*

*CGGTAATCCATGCCGAGTAGCGTAGATCCATGCCGATTACCGTAACTATCCATGCCGATCCAGCGTAG*
*GCCATTAGGTACGGCTCATCGCATCTAGGTACGGCTAATGGCATTGATAGGTACGGCTAGGTCGCATC*

*CTAATCCATGCCGATCCATCAGTACCGATTATCCATGCCGATCCATGCACGCAGGAA3'*
*GATTAGGTACGGCTAGGTAGTCATGGCTAATAGGTACGGCTAGGTACGTGCGTCCTT5'*

For *W₉* → *too light*:
**The length of the amplified sequence is 10 bp.**

*5'GCTATGGCTG3'*
*3'CGATACCGAC5'*

For *S₉ → clear*:
**The length of the amplified sequence is 10 bp.**
*5'TCTCTATGGA3'*
*3'AGAGATACCT5'*

*Rule 10:*

For *Td₁₀ → moderate*:
**The length of the amplified sequence is 130 bp.**
*5'CTGGACTGGAATCCATGCCGTAATTTAATTATCCATGCCGATGATCCGATCCATCCATGCCGATCC*
*3'GACCTGACCTTAGGTACGGCATTAAATTAATAGGTACGGCTACTAGGCTAGGTAGGTACGGCTAGG*

*ATTTTATTTTATCCATGCCGATCCATTCAGCTCAGCATCCATGCCGATCCATGCCGAATCGAAT3'*
*TAAAATAAAATAGGTACGGCTAGGAGTCGAGTCGTAGGTACGGCTAGGTACGGCTTAGCTTA5'*

For *ΔT₁₀ → very saturated*:
**The length of the amplified sequence is 10 bp.**
*5'TTCGTTTCGT3'*
*3'AAGCAAAGCA5'*

For *ΔT₁₀' → saturating*:
**The length of the amplified sequence is 10 bp.**
*5'GTAACGTAAC3'*
*3'CATTGCATTG5'*

For *W₁₀ → excellent*:
**The length of the amplified sequence is 130 bp.**
*5'GCTATGGCTGATCCATGCCGATACCGATAGATCCATGCCGATTAACCCGTGGATCCATGCCGATCCC*
*3'CGATACCGACTAGGTACGGCTATGGCTATCTAGGTACGGCTAATTGGGCACCTAGGTACGGCTAGGG*

*AGTAAGTCTATCCATGCCGATCCATATTCCGATGCATCCATGCCGATCCATGCTACAGTACGG3'*
*TCATTCAGATAGGTACGGCTAGGTATAAGGCTACGTAGGTACGGCTAGGTACGATGTCATGCC5'*

For *S₁₀ → clear*:
**The length of the amplified sequence is 10 bp.**
*5'TCTCTATGGA3'*
*3'AGAGATACCT5'*

*For Fuzzy DNA of Td = 25°C (i.e. Td_ob)*
**The length of the amplified sequence is 210 bp.**
*5'CTGGACTGGATAATTTAATTGATCCGATCCATATTTTATTTTATCCTCAGCTCAGCATCCATCGAA*
*3'GACCTGACCTATTAAATTAACTAGGCTAGGTATAAAATAAAATAGGAGTCGAGTCGTAGGTAGCTT*

*TCGAATATCCATGCAATGTAATGTATCCATGCCGCCAGACCAGAATCCATGCCGATATCGTATCGTAT*
*AGCTTATAGGTACGTTACATTACATAGGTACGGCGGTCTGGTCTTAGGTACGGCTATAGCATAGCATA*

*CCATGCCGATCCTTAGATTAGAATCCATGCCGATCCATATTCAATTCAATCCATGCCGATCCATGCGC*
*GGTACGGCTAGGAATCTAATCTTAGGTACGGCTAGGTATAAGTTAAGTTAGGTACGGCTAGGTACGCG*

*CAAGCCAA3'*

*GTTCGGTT5'*

*Fuzzy DNA of ΔT = -4.0°C (i.e. ΔT$_{ob}$)*
**The length of the amplified sequence is 110 bp.**
*5'TTCGTTTCGTATCCATGCCGATCAAACCAAACATCCATGCCGATCCCGGAACGGAAATCCATGCCGA*
*3'AAGCAAAGCATAGGTACGGCTAGTTTGGTTTGTAGGTACGGCTAGGGCCTTGCCTTTAGGTACGGCT*

*TCCATATCCGATCCGATCCATGCCGATCCATGCGAATGGAATG3'*
*AGGTATAGGCTAGGCTAGGTACGGCTAGGTACGCTTACCTTAC5'*

*Fuzzy DNA of ΔT ' = -2.0°C (i.e. ΔT'$_{ob}$)*
**The length of the amplified sequence is 88 bp.**
*5'GTAACGTAACATCCATGCCGATCCAAATAAAATAATCCATGCCGATCCATTGCGATGATTATCCATG*
*3'CATTGCATTGTAGGTACGGCTAGGTTTATTTTATTAGGTACGGCTAGGTAACGCTACTAATAGGTAC*

*CCGATCCATGCCGGAATTAAC3'*
*GGCTAGGTACGGCCTTAATTG5'*

*Fuzzy DNA of W = 7.0Knts/hr. (i.e. W$_{ob}$)*
**The length of the amplified sequence is 110 bp.**
*5'GCTATGGCTGATCCATGCCGATATACCGATAGATCCATGCCGATCCTAACCCGTGGATCCATGCCGA*
*3'CGATACCGACTAGGTACGGCTATATGGCTATCTAGGTACGGCTAGGATTGGGCACCTAGGTACGGCT*

*TCCATCAGTAAGTCTATCCATGCCGATCCATGCATTCCGATGC3'*
*AGGTAGTCATTCAGATAGGTACGGCTAGGTACGTAAGGCTACG5'*

*Fuzzy DNA of S = 10.1% (i.e. S$_{ob}$)*
**The length of the amplified sequence is 38 bp.**
*5'TCTCTATGGAATCCATGCCGATCCATGCCGATAGCAAT3'*
*3'AGAGATACCTTAGGTACGGCTAGGTACGGCTATCGTTA5'*

**b.** *The amplified sequences are made run through the agarose gel to separate the sequences according to their length by gel electrophoresis. The sequences of each domain are made run in separate gel (or different lanes in same gel) to identify which sequences in the antecedent parts of the rules are similar (i.e. of minimum difference in length) to the given problem.*

We have five gels for each of the five domain.

Table 9.  Length differences (in bp) with observed data after amplification

| Length of Sequences | Td | length difference | ΔT | length difference | ΔT ' | length difference | W | length difference | S | length difference | average dissimilarity |
|---|---|---|---|---|---|---|---|---|---|---|---|
| Rule 1: | 10 | 200 | 10 | 100 | 192 | 104 | 10 | 100 | 182 | 144 | 129.6 |
| Rule 2: | 10 | 200 | 146 | 36 | 10 | 78 | 130 | 20 | 10 | 28 | 72.4 |
| Rule 3: | 130 | 80 | 202 | 92 | 192 | 104 | 202 | 92 | 110 | 72 | 88 |
| Rule 4: | 158 | 52 | 146 | 36 | 192 | 104 | 130 | 20 | 10 | 28 | 48 |
| Rule 5: | 158 | 52 | 202 | 92 | 10 | 78 | 202 | 92 | 182 | 144 | 91.6 |
| Rule 6: | 202 | 8 | 10 | 100 | 10 | 78 | 10 | 100 | 10 | 28 | 62.8 |
| Rule 7: | 202 | 8 | 146 | 36 | 192 | 104 | 202 | 92 | 110 | 72 | 62.4 |

| | | | | | | | | | | |
|---|---|---|---|---|---|---|---|---|---|---|
| Rule 8: | 130 | 80 | 146 | 36 | 10 | 78 | 130 | 20 | 182 | 144 | 71.6 |
| Rule 9: | 130 | 80 | 10 | 100 | 192 | 104 | 10 | 100 | 10 | 28 | 82.4 |
| Rule 10: | 130 | 80 | 10 | 100 | 10 | 78 | 130 | 20 | 10 | 28 | 61.2 |
| Observed data: | 210 | - | 110 | - | 88 | - | 110 | - | 38 | - | - |

*c. The sequences representing the consequent parts are all known from the given data. These consequences have short DNA sequences representing the elements of the corresponding domains. The membership values of the elements of the consequent domain are reduced depending on the average of dissimilarity among the antecedent clauses. Here the threshold of the difference between lengths of fuzzy DNA sequences has to be suitably chosen depending on the need of the problem.*

Table 10. Reduction in membership value (bases) due to length difference of the domain

| Length difference | Reduction in membership value in terms of bases |
|---|---|
| 1 to 15 | 1 |
| 16 to 30 | 2 |
| 31 to 55 | 4 |
| 56 to 70 | 8 |
| 71 to 85 | 12 |
| 86 to 100 | 16 |
| above 100 | 20 |

The threshold value of similarity measurement is set to be 30 bp. If the length difference is more than 30 bp for all domains, then the rule is not accepted.

By inspecting the gels of these two domains we can conclude that, the amplified sequences of $Td_6$ and $Td_7$ are close to the amplified fuzzy DNA of Td' (difference in length is 8 bp). The amplified sequences of $W_2$, $W_4$, $W_8$ and $W_{10}$ are close to the amplified fuzzy DNA of W' (difference in length is 20 bp). The amplified sequences of $S_2$, $S_4$, $S_6$, $S_9$ and $S_{10}$ are close to the amplified fuzzy DNA of S' (difference in length is 28 bp).

So, we can derive our resultant sequence from the consequences of the rules whose antecedence are close to the observed data. The selected consequences are given below:

$C_2$ :
5'CTACCATCAGATCCATGCCGTCGGTGATGGATCCATGCCGATCCGAGCATAGGGATCCATGCCGATC
3'GATGGTAGTCTAGGTACGGCAGCCACTACCTAGGTACGGCTAGGCTCGTATCCCTAGGTACGGCTAG

CATGCCGCTAACTTGTAATCCATGCCGATCCGTCTGATACGATCCATGCCG3'
GTACGGCGATTGAACATTAGGTACGGCTAGGCAGACTATGCTAGGTACGGC5'

$C_4$ :
5'CTACCATCAGATCCATGCCGATCCTCGGTGATGGATCCATGCCGATCCATGCCGGAGCATAGGGATC
3'GATGGTAGTCTAGGTACGGCTAGGAGCCACTACCTAGGTACGGCTAGGTACGGCCTCGTATCCCTAG

                      CATGCCGATCCCTAACTTGTAATCCATGCCGGTCTGATACGATCCAT3'
                      GTACGGCTAGGGATTGAACATTAGGTACGGCCAGACTATGCTAGGTA5'

$C_6$ :
5'CTACCATCAGATCCATGCCGATCCTCGGTGATGGATCCATGCCGATCCATGCCGGAGCATAGGGATC
3'GATGGTAGTCTAGGTACGGCTAGGAGCCACTACCTAGGTACGGCTAGGTACGGCCTCGTATCCCTAG

                      CATGCCGATCCCTAACTTGTAATCCATGCCGGTCTGATACGATCCAT3'
                      GTACGGCTAGGGATTGAACATTAGGTACGGCCAGACTATGCTAGGTA5'

$C_7$ :
5'CTACCATCAGATTCGGTGATGGATCCATGAGCATAGGGATCCATGCCGCTAACTTGTAATCCATGCC
3'GATGGTAGTCTAAGCCACTACCTAGGTACTCGTATCCCTAGGTACGGCGATTGAACATTAGGTACGG

                      GATCCGTCTGATACGATCCATGCCGATCCATGCCG3'
                      CTAGGCAGACTATGCTAGGTACGGCTAGGTACGGC5'

$C_8$ :
5'CTACCATCAGATCCATTCGGTGATGGATCCATGCCGGAGCATAGGGATCCATGCCGATCCCTAACTT
3'GATGGTAGTCTAGGTAAGCCACTACCTAGGTACGGCCTCGTATCCCTAGGTACGGCTAGGGATTGAA

                      GTAATCCATGCCGATCCATGCCGGTCTGATACGATCCATGCCGATCC3'
                      CATTAGGTACGGCTAGGTACGGCCAGACTATGCTAGGTACGGCTAGG5'

$C_9$ :
5'CTACCATCAGATCCATGCCGTCGGTGATGGATCCATGCCGATCCGAGCATAGGGATCCATGCCGATC
3'GATGGTAGTCTAGGTACGGCAGCCACTACCTAGGTACGGCTAGGCTCGTATCCCTAGGTACGGCTAG

                      CATGCCGCTAACTTGTAATCCATGCCGATCCGTCTGATACGATCCATGCCG3'
                      GTACGGCGATTGAACATTAGGTACGGCTAGGCAGACTATGCTAGGTACGGC5'

$C_{10}$ :
5'CTACCATCAGATCCATGCCGATCCATGCCGTCGGTGATGGATCCATGCCGATCCGAGCATAGGGATC
3'GATGGTAGTCTAGGTACGGCTAGGTACGGCAGCCACTACCTAGGTACGGCTAGGCTCGTATCCCTAG

                      CATGCCGCTAACTTGTAATCCATGTCTGATACGAT3'
                      GTACGGCGATTGAACATTAGGTACAGACTATGCTA5'

- For $C_2$, the membership values of the elements are reduced by 12 bases as the average dissimilarity is 72 (approx.) bp.
- For $C_4$, the membership values of the elements are reduced by 4 bases as the average dissimilarity is 48 bp.

- For $C_6$, the membership values of the elements are reduced by 8 bases as the average dissimilarity is 63 (approx.) bp.
- For $C_7$, the membership values of the elements are reduced by 8 bases as the average dissimilarity is 62 (approx.) bp.
- For $C_8$, the membership values of the elements are reduced by 12 bases as the average dissimilarity is 72 (approx.) bp.
- For $C_9$, the membership values of the elements are reduced by 12 bases as the average dissimilarity is 82 (approx.) bp.
- For $C_{10}$, the membership values of the elements are reduced by 8 bases as the average dissimilarity is 61 (approx.) bp.

From the given data the all these DNA sequences of the consequences are known. The short DNA oligonucleotide sequences representing the elements of $C_2$, $C_4$, $C_6$, $C_7$, $C_8$, $C_9$ and $C_{10}$ along with their reduced membership values are taken in a test tube.

For $C_2$ membership value reduced by 12 bases. The short sequences are
5'CTACCATCAG3', 5'TCGGTGATGGAT3', 5'GAGCATAGGGATCCATGC3', 5'CTAACTTGTAAT3' and 5'GTCTGATACG3'.

For $C_4$ membership value reduced by 4 bases. The short sequences are
5'CTACCATCAGATCCATGCCG3', 5'TCGGTGATGGATCCATGCCGATCCAT3', 5'GAGCATAGGGATCCATGCCG3', 5'CTAACTTGTAATCCAT3' and 5'GTCTGATACGAT3'.

For $C_6$ membership value reduced by 8 bases. The short sequences are
5'CTACCATCAGATCCAT3', 5'TCGGTGATGGATCCATGCCGAT3', 5'GAGCATAGGGATCCAT3', 5'CTAACTTGTAAT3' and 5'GTCTGATACG3'.

For $C_7$ membership value reduced by 8 bases. The short sequences are
5'CTACCATCAG3', 5'TCGGTGATGG3', 5'GAGCATAGGGAT3', 5'CTAACTTGTAATCCAT3' and 5'GTCTGATACGATCCATGCCGAT3'.

For $C_8$ membership value reduced by 12 bases. The short sequences are
5'CTACCATCAG3', 5'TCGGTGATGG3', 5'GAGCATAGGGAT3', 5'CTAACTTGTAATCCATGC3' and 5'GTCTGATACGAT3'.

For $C_9$ membership value reduced by 12 bases. The short sequences are
5'CTACCATCAG3', 5'TCGGTGATGGAT3', 5'GAGCATAGGGATCCATGC3', 5'CTAACTTGTAAT3' and 5'GTCTGATACG3'.

For $C_{10}$ membership value reduced by 8 bases. The short sequences are 5'CTACCATCAGATCCATGCCGAT3', 5'TCGGTGATGGATCCAT3', 5'GAGCATAGGGAT3', 5'CTAACTTGTA3' and 5'GTCTGATACG3'.

Step 3 of the algorithm:

**a.** *The single stranded short sequences are made double stranded to perform the process of sequence specific DNA isolation.*

**b.** *Affinity purification is conducted with the complementary sequences to each DNA oligonucleotide representing the elements of the consequent part. The sequences having identical short sequences are stored in the same test tube.*

The sequences used for affinity purification are: 3'GATGGTAGTC5' (complementary to 5'CTACCATCAG3'), 3'AGCCACTACC5' (complementary to 5'TCGGTGATGG3'), 3'CTCGTATCCC5' (complementary to 5'GAGCATAGGG3'), 3'GATTGAACAT5' (complementary to 5'CTAACTTGTA3') and 3'CAGACTATGC5' (complementary to 5'GTCTGATACG3'). This step is done to separate all the short sequences having identical oligonucleotides representing each element of the consequence. Now we have five test tubes.

Sequences of test tube 1: 5'CTACCATCAG3', 5'CTACCATCAGATCCATGCCG3', 5'CTACCATCAGATCCAT3', 5'CTACCATCAG3' and 5'CTACCATCAGATCCATGCCGAT3'.

Sequences of test tube 2: 5'TCGGTGATGGAT3', 5'TCGGTGATGGATCCATGCCGATCCAT3', 5'TCGGTGATGGATCCATGCCGAT3', 5'TCGGTGATGG3' and 5'TCGGTGATGGATCCAT3'

Sequences of test tube 3: 5'GAGCATAGGGATCCATGC3', 5'GAGCATAGGGATCCATGCCG3', 5'GAGCATAGGGATCCAT3'and 5'GAGCATAGGGAT3'.

Sequences of test tube 4: 5'CTAACTTGTAAT3', 5'CTAACTTGTAATCCAT3', 5'CTAACTTGTAATCCATGC3' and 5'CTAACTTGTA3'.

Sequences of test tube 5: 5'GTCTGATACG3', 5'GTCTGATACGAT3' and 5'GTCTGATACGATCCATGCCGAT3'.

**c.** *Gel electrophoresis is performed with the sequences of each test tubes separately. This step is done to get the sequences having highest length (i.e. the sequence having highest membership value attached to it). After gel electrophoresis the sequences having highest length are extracted from each gel.*

The DNA short sequence having highest length is extracted from the content of each test tube by performing gel electrophoresis. The extracted sequences are:

- 5'CTACCATCAGATCCATGCCGAT3'
  3'GATGGTAGTCTAGGTACGGCTA5'

- 5'TCGGTGATGGATCCATGCCGATCCAT3'
  3'AGCCACTACCTAGGTACGGCTAGGTA5'

- 5'GAGCATAGGGATCCATGCCG3'
  3'CTCGTATCCCTAGGTACGGC5'

- 5'CTAACTTGTAATCCATGC3'
  3'GATTGAACATTAGGTACG5'

- 5'GTCTGATACGATCCATGCCGAT3'
  3'CAGACTATGCTAGGTACGGCTA5'

**d.** *The blunt ends of the double stranded DNA sequences selected from the previous step are ligated by T4 DNA ligase. The resultant sequences after ligation have different lengths.*

Example:
5'CTACCATCAGATCCATGCCGATGAGCATAGGGATCCATGCCGCTAACTTGTAATCCATGC3'
3'GATGGTAGTCTAGGTACGGCTACTCGTATCCCTAGGTACGGCGATTGAACATTAGGTACG5'

5'TCGGTGATGGATCCATGCCGATCCATCTAACTTGTAATCCATGC3'
3'AGCCACTACCTAGGTACGGCTAGGTAGATTGAACATTAGGTACG5'

5'GTCTGATACGATCCATGCCGATCTAACTTGTAATCCATGCTCGGTGATGGATCCATGCCGATCCAT
3'CAGACTATGCTAGGTACGGCTAGATTGAACATTAGGTACGAGCCACTACCTAGGTACGGCTAGGTA

CTACCATCAGATCCATGCCGATGTCTGATACGATCCATGCCGATGAGCATAGGGATCCATGCCG3'
GATGGTAGTCTAGGTACGGCTACAGACTATGCTAGGTACGGCTACTCGTATCCCTAGGTACGGC5'

etc.

**e.** *To get the resultant sequence the following steps are performed:*

  i. *Apply the method of affinity purification to make sure that in every sequence each short sequences are present at least once.*
  ii. *Perform gel electrophoresis by the above sequences of different length to extract the sequences having the length 90 to 120 bp.*
  iii. *The selected sequence from the previous step is the most probable result of the problem. The order of the bases of the resultant sequence can be obtained from sequencer.*

## 5.5. Result

The final consequence i.e. the double stranded DNA sequence representing the possibility of visibility is:

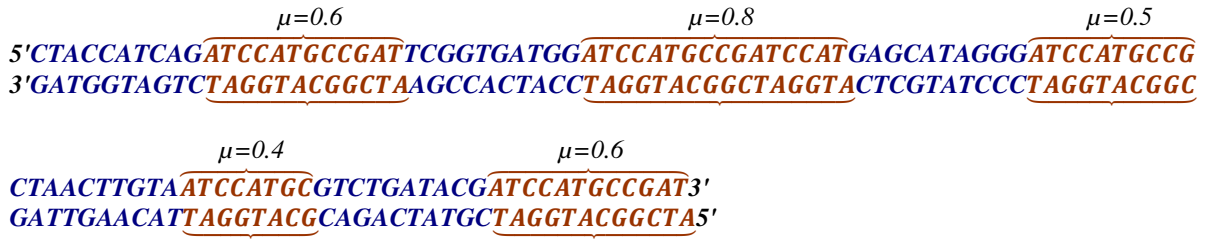

In this experiment we have encoded the expert rules in form of DNA sequence. In the problem five parameters of radiation fog (i.e. Dew Point, Dew point Spread, The rate of change of Dew point Spread per day, Wind Speed and Sky Condition) are given. We have to predict the radiation fog i.e. the possibility of visibility.

By inspecting the sequence obtained from sequencer, the membership value of each primary fuzzy sets of visibility domain is given below:

Membership value of the sequence representing **very low**  → **0.6**
Membership value of the sequence representing **low**       → **0.8**
Membership value of the sequence representing **medium**    → **0.5**
Membership value of the sequence representing **high**      → **0.4**
Membership value of the sequence representing **very high** → **0.6**

Therefore, the DNA sequence TCGGTGATGG representing the particular primary fuzzy set i.e. low has the highest membership value 0.8.

The resultant sequence has been derived from the consequences of the IF-THEN rules depending on the similarity among the antecedent parts. Thus, the resultant sequence reflects the influences of the expert rules.

## 6. Conclusion

      We propose fuzzy reasoning based on DNA computing and obtained satisfactory results at the end of our wet lab experiment. We verify the performance of our wet lab algorithm in terms of prediction of radiation fog. We get very satisfactory performance in our method using wet lab algorithm. Fuzzy reasoning based on DNA computing simply enhances the existing tool of fuzzy reasoning by bringing it down to nanoscale. The DNA chemistry of fuzzy reasoning can avoid the tedious choice of a suitable implication operator which is absolutely necessary for fuzzy logic based fuzzy reasoning. We consider the Fuzzy DNA to handle the vague concept of day to day reasoning which is very prominent in all real life applications. We have generated a new measure of similarity which is different from the existing methods of similarity measures (Yeung and Tsang, 1997; 1998) and applied to the present form of applicable fuzzy reasoning (see section 5.3). This new method is very much suitable for wet lab implementation. Like other DNA computing models based on DNA operations we have also achieved successful result in the wet lab implementation. We can apply the present approach to fuzzy reasoning based on DNA computing to different areas of pattern classification, object recognition, control problems, weather forecasting etc.

**References**

Adleman, L. (1994), "Molecular computation of solutions to combinatorial problems", Science, Vol. 266, pp. 1021-1024.

Adleman, L. "On constructing a molecular computer." ftp://usc.edu/pub/csinfo/papers/adleman

Adleman, L., Rothemund, P., Roweis, S., Winfree, E. (1996), "On applying molecular computation to the Data Encryption Standard. $2^{nd}$ DIMACS workshop on DNA based computers", Princeton, 28-48.

Bach E., Condon A., Glaser E. and Tanguay C., (1996), "DNA models and algorithms for NP-complete problems." March 1996.

Beaver, D. (1995), "Computing with DNA." Journal of Computational Biology, (2:1).

Boneh D., Lipton R., Sgall J., Dunworth C., "Making DNA computers error resistant", 2nd DIMACS workshop on DNA based computers, Princeton, 1996, pp. 102-110.

Chang, Weng-Long. Ho, Michael. and Guo, Minyi. (2004), "Molecular Solutions for the Subset-sum Problem on DNA-based Supercomputing", BioSystems (Elsevier Science), Vol. 73, No. 2, pp. 117-130.

Chang W.L. and Guo M., (2003)"Solving the Set-cover Problem of Exact Cover by 3-Sets in the Adleman-Lioton's Model", Biosystems, Vol. 72, No.3, pp.263-275, 2003.

Chen S.M., Yeh M.S., Hsiao P.Y., "A comparison of Similarity Measures of Fuzzy Values", in Fuzzy Sets and System 72 (1995) 79-89.

Hartmanis J., On the weight of computations. Bulletin of the European Association of theoretical Computer Science, 55: (1995), 136-138.

Head, T., (1987), "Formal language theory and DNA: an analysis of the generative capacity of recombinant behaviors", Bulletin of Mathematical Biology, Vol. 49, pp. 737-759.

Head, T.,(1998) "Hamiltonian paths and Double Stranded DNA" – in computing with Bio-molecules; Theory and Experiments, eds: Gheorghe Paun, Springer, 1998, PP 80-92.

Hug, H. and Schuler, R. (2001), "DNA-based parallel computation of simple arithmetic", In Proceedings of the 7th International Meeting on DNA Based Computers, pp. 159-166.

Ibrahim Z., Rose J.A., Tsuboi Y., Ono O., (2006) "A new Readout Approach in DNA Computing Based on Real-Time PCR with TaqMan Probes", Conference Paper, Dec 18, 2006.

Johanyak C., Kovacs S., "Distance Based Similarity Measures of Fuzzy Sets", in SAMI Conference 2005.

Jyh-Fu Jeng D., Watada J., Berlin Wu, Jui-Yu Wu. (2006), "Fuzzy Forecasting with DNA Computing", conference paper, Dec 18, 2006.

Kari L., (1997) DNA computing – The arrival of biological mathematics. The Mathematical Intelligencer, 19(1997), 9–22.\

Kurtz, S. Mahaney, J. Royer, J. Simon, Active transport in biological computing. 2nd DIMACS workshop on DNA based computers, Princeton, (1996), 111-121.

Lipton, R., DNA solution of hard computational problems. Science, 268: (April, 1995), 542-545.

Mizumoto, M. (1985a) : Extended fuzzy reassoning, Approximate Reasoning in Expert Systems (ed. Gupta, M.M., Kandel, A., Bandler, W. & Kiszka, J.B.), Elsevier Science Publishers B.V.(North-Holland), 71-85 (1985).

Mizumoto,M. (1985b) : Fuzzy reasoning for "If...Then...Else..." under new compositional rules of inference, Management Decision Support Systems Using Fuzzy Sets and Possibility Theory (ed. by J.Kacprzyk and R.R.Yager), Verlag TUV Rheinland, W.Germany, 229-239 (1985).

Ray K.S. and Chatterjee P.,(2008) "Approximate Reasoning Based on DNA Computing" (Technical Report) TR ECSU no: 2/08 of Indian Statistical Institute, Kolkata.

Ray K.S. and Ghoshal J., 2000 K.S. Ray and J. Ghoshal, "Neuro-genetic approach to multidimensional fuzzy reasoning for pattern classification" , Fuzzy Sets and Systems, Volume 112, Issue 3, pages 449-483.

Reif J.H. and LaBean T.H. (2009) DNA Nanotechnology and its Biological Applications. In book "Bio-inspired and Nanoscale Integrated Computing", Chapter 13, pp. 349-375 (edited by Mary Mehrnoosh Eshaghian-Wilner), Publisher: Wiley, Hoboken, NJ, USA, (February 2009).

Reif J.H. and LaBean T.H., (2009) Engineering Natural Computation by Autonomous DNA-Based Biomolecular Devices, Invited Chapter, Handbook of Natural Computing, edited by Grzegor Rozenberg, Thomas Bck, Joost Kok, Springer-Verlag, New York.

Yeung D.S. and Tsang E. C. C., (1997), "A comparative study on Similarity based fuzzy reasoning method", IEEE Trans Syst.Man, Cyber.B, Vol 27, 1997, PP 16-227.

Yeung D.S. and Tsang E. C. C., "A Multileveled Weighted Fuzzy Reasoning Algorithm for Expert Systems", IEEE TRANSACTIONS ON SYSTEMS, MAN, AND CYBERNETICS-PART A: SYSTEMS AND HUMANS, VOL. 28, NO. 2, MARCH 1998.

Zadeh L.A., (1970) "Theory of Approximate Reasoning" in: J. E. Hayes, Donald Michie and L. I. Mukulich, Ed., Machine Intelligence (Ellis Herwood), 1970, PP 144-194.